\documentstyle[12pt]{article}
\newcommand{\beq}{\begin{eqnarray}}
\newcommand{\eeq}{\end{eqnarray}}

\newcommand{\pr}{\partial}

\newcommand{\rf}[1]{(\ref{#1})}

\renewcommand{\l}{\lambda}

\renewcommand{\b}{\beta}
\renewcommand{\a}{\alpha}

\newcommand{\eps}{\epsilon}

\newcommand{\zbar}{{\bar z}}
\newcommand{\Abar}{{\bar A}}
\newcommand{\Qbar}{{\bar Q}}
\newcommand{\Ubar}{{\bar U}}
\newcommand{\Vbar}{{\bar V}}
\newcommand{\Wbar}{{\bar W}}
\newcommand{\B}{{\cal B}}

\newcommand{\Bbar}{{{\bar{\cal B}}}}

\begin{document}

\topmargin 0pt
\oddsidemargin 5mm
\headheight 0pt

\topskip 5mm


\thispagestyle{empty}
      
\begin{flushright}
BCUNY-HEP-99-3\\
December 1999\\
Revised, March 1999
\hfill
\end{flushright}

\begin{center}

\hspace{10cm}

\vspace{15pt}
{\large \bf
EXTREMAL CURVES IN $2+1$-DIMENSIONAL YANG-MILLS THEORY}

\vspace{20pt}

{\bf Peter Orland}\footnote{\noindent Work supported 
by PSC-CUNY Research Award Program 
grants
nos. 668460, 61508-00 and 69466-00 and CUNY Collaborative
Incentive Grant 91915-00-06.}

Center for Theoretical Physics,\\
The Graduate School and University Center,\\
The City University of New York\\ and\\
Baruch College, The City University of New York\\
New York, NY, U.S.A.\\
orland@gursey.baruch.cuny.edu
\\

\vspace{10pt}
and
\vspace{10pt}

{\bf Gordon W. Semenoff}\footnote{\noindent Work supported
by NSERC of Canada and the Niels Bohr Fund of Denmark. }\\

Department of Physics and Astronomy,\\
University of British Columbia,\\
6224 Agricultural Road\\
Vancouver, British Columbia, Canada V6T 1Z1\\
semenoff@physics.ubc.ca

\vspace{30pt}

{\bf Abstract}

\end{center} 

We examine the structure of the potential energy of 2+1-dimensional
Yang-Mills theory on a torus with gauge group SU(2). We use
a standard definition of
distance on the space of gauge orbits.  A curve of extremal
potential energy in orbit space defines connections
satisfying a certain
partial differential equation. We argue that the energy
spectrum
is gapped because the extremal
curves are of finite length. Though classical
gluon waves
satisfy our differential equation, they are not extremal
curves. We construct examples of extremal curves and
find how the length of
these curves
depends on the dimensions of the torus. The intersections
with the Gribov horizon are determined
explicitly.  The results are discussed in the context of Feynman's ideas
about the origin of the mass gap.

\vfill
\newpage
\pagestyle{plain}
\setcounter{page}{1}
\setcounter{footnote}{0}

\section{Introduction}

\setcounter{equation}{0}
\renewcommand{\theequation}{1.\arabic{equation}}

To understand how the non-Abelian structure of gauge theories can lead to
confinement and a dynamical mass gap is a classic problem. In this paper,
we examine the geometry of the gauge-theoretic field-configuration space
in two space and one time dimension and
the potential-energy functional on this space. We examine the
region of this infinite-dimensional space where the potential energy is
small (i.e. of order one over the size of the system). A major
question is whether this region is of finite extent. In this
paper, we partially answer this question. We discuss
how our results support the hypothesis that the shape of the
potential-energy functional and the geometry of configuration space lead
to the presence of a mass gap when the theory is quantized. 

Early attempts to understand confinement and the gap in terms of the
properties of the configuration space (or {\it orbit space})
were made by Gribov \cite{gribov2}, Feynman \cite{feynman} and Singer
\cite{singer2}. Gribov and Feynman attempted to use gauge transformations
to minimize the Pythagorean
distance between configurations. Feynman, in particular made certain
conjectures concerning the mass spectrum and confinement in $2+1$
dimensions\footnote{We refer only to Feynman's published article
\cite{feynman}
and not
to the preprint on the subject written earlier by him.}. The first
general use of this minimal distance in the
literature may have been in reference \cite{ahs}.

Singer defined a
metric of Riemannian form on orbit
space, for the purpose of studying the spectrum of
the Laplacian on this space
in a gauge-invariant way \cite{singer2}. Singer's
metric is the infinitesimal version of that discussed in references
\cite{ahs}
and \cite{feynman}. Further discussion
of the metric
can be found in \cite{miscellaneous} and in particular \cite{bv1}. Other
important observations
were made by Zwanziger, van Baal and Cutkowsky \cite{ZVC}. A recent
related development is the important work in $2+1$ dimensions of Karabali
and Nair who solved Gauss' law and have demonstrated the existence of
confinement and a mass gap at strong coupling with no cut-off \cite{kn1} and
developed strong-coupling expansions which seem quite reliable \cite{kn2}. 

The degrees of freedom of the gauge theory are the set of gauge
connections modulo gauge transformations.  These degrees of freedom are
called gauge orbits. The geometry of the space of gauge orbits (called
orbit space) is known to be quite complicated
\cite{singer2,bv1,gribov1,singer1,orland}. It was shown in
reference \cite{orland} that the distance between configurations is a
metric (in the real-analytic sense) and is just the length of a minimal
geodesic with the local metric of Singer, as had been conjectured by
Babelon and Viallet \cite{bv1}. Using the lattice formulation of gauge
theory, it was first shown in reference \cite{kmo} and
later in reference \cite{becchi} that the heat
kernel of the kinetic term of the
Hamiltonian is proportional to the exponential of the square of this
metric.

Our starting point on this problem is the intuitive picture of the
dynamics of $2+1$-dimensional gauge theory, proposed by Feynman
\cite{feynman}. Feynman worked in the Schr\"{o}dinger representation and
analyzed the structure of the potential energy functional.  We interpret
Feynman's approach \cite{feynman} as the study of the dependence of the
potential energy on a coordinate called the radius. This
coordinate is the distance
from the pure gauge orbit. Feynman argued that, unlike the case of
Abelian gauge theories, all
non-Abelian orbits with minimal potential
energy occur within a region of finite radius (incidentally, this
conjecture is false for Yang-Mills
theories in three space dimensions \cite{orland,kmo} and the O(n)
nonlinear sigma
model in one space dimension \cite{kmo,moreno}). Feynman gave some
heuristic
arguments to this effect. We construct extrema of the potential energy on
spheres of fixed distance from the pure gauge orbit in orbit space
for the SU(2) gauge theory.  We
find a differential equation which must be satisfied by these extrema.  On
a case-by-case basis, we determine which of the solutions of this equation
are truly extrema and call these solutions {\it extremal
curves} \cite{moreno}. We also
show that our extremal curves
are actually minimal curves or {\it river valleys} \cite{kmo,moreno}. The
only extremal curves we have succeeded in finding
have finite length in orbit space. If all the
extremal curves of the $2+1$-dimensional SU(2)
Yang-Mills theory are of
finite length, then we expect a gap in the energy spectrum directly
above the ground
state.

While we have not yet
constructed all the extremal curves, we have found and analyzed two
interesting subclasses when space is a flat torus: \begin{enumerate}
\item Flat families of orbits. These are nontrivial zero-curvature orbits. 
\item A family of ``special" non-Abelian orbits. These are proved to be the
only extremal-curve orbits with constant potential-energy density. 
\end{enumerate}
We prove that 1. and 2. have a fixed
maximum radius and are river valleys.

We also consider some
``Abelian'' families of orbits.  These contain
representative gauge fields
in a U(1) subalgebra of the SU(2) theory that we consider. These are
standing color waves, i.e. the classical analogues of gluons. We show
that these families of orbits are not extremal curves. This result
is important because it strongly indicates that gluons are not physical
excitations. In contrast, the river valleys of an Abelian gauge
theory are precisely such standing waves; this is why the physical
excitations of this theory
are photons.

There is a subtlety in this analysis which has to do with the fact that
gauge connections lie in the adjoint representation of the Lie algebra. 
The true gauge group of pure Yang-Mills theory (in the continuum) is
${\rm SU(2)}/Z_2\simeq {\rm SO(3)}$, rather than SU(2).  In fact, we shall have a
choice whether to use SO(3) or SU(2) as the gauge group.  The difference
is important when the space has non-contractable loops and must be 
considered when periodic boundary conditions are imposed. 
 
In the next section we briefly review the motivation for
the metric we use on orbit space. In Section 3 we show how extremal
curves are solutions of a nonlinear hyperbolic or parabolic
differential equation, and discuss
how to distinguish between river valleys, other extremal curves and
unphysical solutions. In section 
4, some general features of the
$2+1$-dimensional SU(2) gauge theory are summarized. Embedded Abelian
curves, are discussed in Section 5; these turn out to be irrelevant. The
simplest river valleys, non-pure-gauge connections of zero curvature
on the torus are investigated in Section 6. We show that the elements
of these river
valleys are within a finite distance of the pure gauge
orbit. In Section 7, we discuss
the solutions of our differential equation with constant potential
energy density (many of the details are presented in the
appendix). We show in Section 8 that these constant-potential-energy
solutions are not extremal curves, except one, which happens to be a
river valley (even this is not an extremal curve unless the volume of space
is finite). The points of this river valley are again with a finite
distance of the pure gauge orbit. In Section 9, we compare our results 
with the arguments made by Feynman \cite{feynman} and briefly
discuss why solutions of the elliptic
case of our
differential equation are of no physical significance. We summarize our
basic results and discuss future directions of this research in Section
10.

\section{The metric on Yang-Mills orbit space}
\setcounter{equation}{0}
\renewcommand{\theequation}{2.\arabic{equation}}

In this section, the dimension of spacetime is $D+1$.
We 
denote the space of connections on a flat $D$-dimensional manifold by
${\cal A}$. A connection is
a Lie-algebra-valued field
$A_{i}(x)$, $i=1,...D$, which can also be written in terms
of a real isovector $A_{i}^{a}(x)$ as
$A_{i}(x)=A_{i}^{a}(x) t^{a}$, where $t^{a}$, $a=1,\dots n$
are the generators of the
Lie group $G$. The structure coefficients are $C^{abc}$, defined by
$[t^{a},t^{b}]=iC^{abc}t^{c}$
Denote the set of local gauge transformations $g(x)$ by ${\cal G}$.
We will sometimes write 
connections leaving the index and space dependence implicit,
e.g. $A$ for $A_{i}(x)$.  The
connection $A$ changes to $A^{g}$ under a local gauge transformation:
\beq
A^{g}_{i}(x)=g^{-1}(x)A_{i}(x)g(x)+ig^{-1}(x)\partial_{i} g(x) \;.
\label{gaugetrans}
\eeq
The covariant derivative is
$D_{i}=\partial_{i}-iA_{i}(x)$.

Orbit space ${\cal O}\equiv{\cal A}/{\cal G}$ is a metric
space \cite{orland}, provided that definitions are made
carefully\footnote{Lebesgue
measure is used, 
gauge fields $A$ are $L^{2}$ functions, the connections $ig^{-1}\partial g$ 
are 
also $L^{2}$ for allowed gauge transformations $g$ and gauge equivalence 
is defined through sequences of gauge transformations. To prove
that gauge equivalence is indeed an equivalence relation and that the
axioms of a metric space and the completeness property
are satisfied takes considerable work. The proofs
are much easier on the lattice \cite{kmo}.}. The
metric is a function of two
variables $\a$ and $\b$ in ${\cal O}$ containing a representative 
connection $A$
and $B$ respectively. Its definition is
\beq
\rho[\a,\b]^{2}= \inf_{g\in {\cal G}} \;\frac{1}{2} \int d^{D}x \; 
{\rm Tr}[A^{g}_{i}(x)-B_{i}(x)]^{2}\;, \label{metric}
\eeq
where the sum on the space index $i$ is implicit.

The potential energy is $U[\alpha]/e_{0}^{2}$, where $e_{0}$ is
the coupling constant
and the
functional $U[\alpha]$ on an orbit $\a$ containing a
representative $A$ is
\beq
U[\a]=\frac{1}{4} \int d^{D}x \; {\rm Tr}\,F_{ij}(x)F_{ij}(x)\;,
\nonumber
\eeq
where $F_{ij}(x)$
is the curvature, or field strength, defined in the usual way as
\beq
F_{ij}(x)=i[D_{i},D_{j}]=
\partial_{i}A_{j}(x)-\partial_{j}A_{i}(x)-i[A_{i}(x),A_{j}(x)] \;.
\label{curvature}
\eeq
The pure gauge
orbit, whose curvature vanishes, will be denoted by $\a_{0}$.

The heat kernel
of the kinetic term
$K$, for short time intervals $\varepsilon$
behaves as \cite{kmo, becchi}
\beq
\left< \beta \left\vert e^{-K \epsilon} \right\vert \alpha \right> \sim
\exp-\frac{1}{2e_{0}^{2}\varepsilon}\; \rho[\beta, \alpha]^{2}  \;. \nonumber
\eeq
This term therefore describes Brownian motion in orbit space
with the metric $\rho$.

\section{Extremal curves in orbit space}
\setcounter{equation}{0}
\renewcommand{\theequation}{3.\arabic{equation}}

We would like to study the region of orbit space ${\cal O}$ where the
potential energy (or magnetic energy) is ``small". To do this we will
attempt to find the orbits $\a$ in ${\cal O}$ such that $U[a]$ is an
extremum on the sphere of fixed $\rho[\a,\a_{0}]=\rho_{0}$. We will then
view $\rho_{0}$ as a coordinate along which this minimum changes. The
result is a hypersurface in ${\cal O}$ which we call an extremal
curve. If the extremum is a minimum, we call the extremal curve a 
river valley \cite{orland, kmo}.


\begin{center}
{\bf 3.1 The Yang-Mills-Proca equation}
\end{center}
\setcounter{equation}{0}
\renewcommand{\theequation}{3.1.\arabic{equation}}

To find extremal curves we vary the functional
\beq
Q[\a]=U[\a]+\lambda (\rho[\a,\a_{0}]^{2}-\rho_{0}^{2})\;, \label{extrem}
\eeq
where the number $\lambda$ is a Lagrange multiplier. Once obtained, it can
be checked whether this curve is a river valley.

At first glance extremizing \rf{extrem} appears intractable. The
constraint implemented by the Lagrange multiplier contains the {\it radius}
$\rho[\a,\a_{0}]$, which from \rf{metric} is
\beq
\rho[\a,\a_{0}]^{2}=\inf_{g\in {\cal G}} \;\frac{1}{2} \int d^{D}x \;{\rm Tr} (A^{g}_{i})^{2}
\label{distance}
\eeq
which, for most connections, is impossible to evaluate (it is this expression
that was first studied by Gribov \cite{gribov2}). However, the situation 
is not hopeless. Any extremum of $Q$ defined in \rf{extrem} is 
also an extremum of the functional $Q_{1}$ of {\it two} fields $A$ and $g$
\beq
Q_{1}[A,g]=\int d^{D} x\;\left[ \frac{1}{4}{\rm Tr} F_{ij}^{2}+\frac{1}{2}\lambda\; {\rm
Tr}(A^{g}_{i})^{2} \right]
-\lambda \rho_{0}^{2} \label{newfunc}
\eeq
(though the converse is not necessarily true - it only would be true 
if the value of $g(x)$ at the extremum were actually the infimum in 
eqn. \rf{distance} ). Extremizing the functional $Q_{1}$ gives a nonlinear 
field equation 
for $A$ and $g$. The problem is still difficult, but a substantial 
simplification over \rf{extrem}. Once an
extremum $\{A,g\}$ of $Q_{1}$ is found, the orbit $\a$ containing the 
resulting connection $A$ can be then be tested 
to see if it is an extremum of $Q$. Suppose we have found an extremum 
$\a$ of $Q_{1}$ defined in
\rf{newfunc}, with representative connection
$A$. Then there is
some $g\in {\cal G}$ such that the radius is given by
\beq
\rho[\a, \a_{0}]^{2} = \frac{1}{2} \int d^{D} x\; {\rm Tr}(A^{g}_{i})^{2}\;. 
\label{squarint}
\eeq
By virtue of the fact that $g$ minimizes the quantity on 
the right-hand side of \rf{squarint}, the
connection $A^{g}$ satisfies the Coulomb gauge condition:
\beq
\partial_{i}A_{i}^{g}=0\;. \label{coulomb}
\eeq
Parenthetically we remark that we ultimately seek $g\in {\cal G}$ 
which is the {\it absolute minimum} of the
right-hand side of \rf{squarint}. But this is a local extremum as 
well, so that \rf{coulomb}
is, in fact, satisfied. The extremal condition for $Q_{1}$ generates 
the equations \rf{coulomb} as well as
\beq
-[D_{i},F_{ij}]+\lambda (A_{j}+i\partial_{j}g\;g^{-1})=0 \;. \label{messyYM}
\eeq
Now redefine $A$ by the gauge-transformed connection $A^{g}$. This is
simply a particular
choice of representative connection of $\a$. Then this choice of $A$
satisfies
\beq
\rho[\alpha, \alpha_{0}]^{2}=\frac{1}{2} \int d^{D}x\; {\rm Tr}
(A_{i})^{2} \;. \label{constraint}
\eeq
We say that the connection $A$ satisfying \rf{constraint} lies in the {\it 
fundamental region} 
\cite{gribov2}, \cite{miscellaneous}, \cite{bv1}, \cite{ZVC}. Furthermore 
the equations \rf{coulomb} and \rf{messyYM} become
\beq
\partial_{i}A_{i}=0\; \label{coulomb1}
\eeq
and
\beq
-[D_{i},F_{ij}]+\lambda A_{j}=0 \;, \label{massiveYM}
\eeq
respectively.

A further simplification is possible. Clearly from the form of
$Q_{1}[A,g]$ in \rf{newfunc}, $Q_{1}[A,g]=Q_{1}[A^{g},1]=Q_{2}[A^{g}]$.
Therefore, one can simply vary $A^{g}$ instead of $A$, in this new
functional $Q_{2}[A^{g}]$. We can now relabel $A^{g}$ by $A$. Therefore,
all that has to be done is to find the extrema of 
\beq Q_{2}[A]=\int d^{D}
x\,\left[ \frac{1}{4}\,{\rm Tr}
 F_{ij}^{2}+\frac{1}{2}\lambda\; {\rm Tr}
\, A_{i}^{2} \right]
-\lambda \rho_{0}^{2}\;. \nonumber
\eeq

At the risk of belaboring the obvious, we will show that the gauge
connection which extremizes $Q_{2}[A]$ automatically obeys the Coulomb
gauge condition \rf{coulomb}, which is a necessary condition for $\int 
{\rm Tr}
(A^{g})^{2} d^{D}x$ to be minimized at $g=1$. In fact, this condition 
follows from
the Proca-Yang-Mills equation \rf{massiveYM} by virtue of
the Jacobi identity. Taking the commutator of $D_{j}$ with both sides of
\rf{massiveYM} yields \beq i[D_{j},[D_{i},[D_{i},D_{j}]]]=\lambda
\partial_{i} A_{i}\;. \label{1}
\eeq
From the Jacobi identity this becomes
\beq
-i[[D_{i},D_{j}],[D_{j},D_{i}]]-i[D_{i},[[D_{i},D_{j}],D_{j}]]=\lambda
\partial_{i}A_{i} \;, \nonumber \eeq which simplifies to \beq
-i[D_{i},[D_{j},[D_{j},D_{i}]]]=\lambda \partial_{i} A_{i}\;. \nonumber
\eeq Notice that this is the same as \rf{1} except the sign of the
left-hand side has changed and \rf{coulomb1} follows. Therefore all of the
extremal curve configurations have a representative connection satisfying
\rf{massiveYM}. We therefore proceed by finding solutions of the
``massive" Yang-Mills or Proca equations
\rf{massiveYM}, then analyzing them to see if they lie on river-valley
gauge orbits. We put ``massive"  in quotes, for as we shall
see in examples below, the ``mass squared" $\lambda$ is negative, so the
equation \rf{massiveYM} is hyperbolic rather than elliptic (the quotes
will be dropped henceforth).

A necessary condition for a family of solutions of the Yang-Mills-Proca 
equation to be an extremal curve is that \rf{constraint} is true. This
means that 
the Faddeev-Popov functional
\beq
{\rm F.P.}=\int d^{D}x\,\, \frac{1}{2}
\left[(\partial_{i}h^{a})^{2}+
C^{abc}A_{i}^{a}\partial_{i}h^{b}h^{c}\right]
\;,\label{faddeev}
\eeq
which is 
the second variation of the integral under a
gauge transformation, is positive.

\begin{center}
{\bf 3.2 Potential-energy stability analysis}
\end{center}
\setcounter{equation}{0}
\renewcommand{\theequation}{3.2.\arabic{equation}}

Suppose we wish to know whether a 
given gauge orbit with a representative connection
in the fundamental region and satisfying the Proca equation is a 
local minimum
of the potential energy. In other words, we seek information as
to whether such a connection lies on a river valley or not. The question 
is similar to that asked in Hamilton-Jacobi
theory, i.e. whether a local extremum of some variational principle is a
local minimum. However, in Hamilton-Jacobi theory it is sufficient to study
the spectrum of linear operators; a luxury which will be denied us.

Suppose that there is a solution of the non-Abelian Proca equation $A$
contained in a gauge orbit $\alpha$, such that $\rho[\alpha, \alpha_{0}]
=\rho_{0}$. Consider
a small variation $A\rightarrow A+\delta A=B$, where
the connection $B$ is contained in the gauge orbit $\beta$. We note that
it is nontrivial to prove that $\rho[\alpha, \beta]$ is then of
order $(\delta A)^{2}$ on the
sphere of radius $\rho_{0}$, i.e. $\rho[\beta, \alpha_{0}]=\rho_{0}$ (this
is done in reference \cite{orland}). We wish to know
whether the new potential energy $U[\beta]$ is greater that
$U[\alpha]$. If for any
variation $\delta A$, we have that $U[\beta]>
U[\alpha]$
then $\alpha$ lies on a river valley.

Let us try to formulate the
statements in the last paragraph more precisely. If both
$\alpha$ and $\beta$
lie on the sphere of radius $\rho_{0}$ about $\alpha_{0}$, then
\beq
\rho_{0}^{2}=\inf_{h} \frac{1}{2} \int d^{D}x\; {\rm Tr} [(A_{i}+\delta 
A_{i})^{h}]^{2}
=\inf_{g} \frac{1}{2} \int d^{D}x \;{\rm Tr} (A_{i}^{g})^{2}
=\frac{1}{2} \int d^{D}x \;{\rm Tr} (A_{i})^{2}\;, \label{sphere}
\eeq
where in the last step, we have assumed that the
connection $A_{i}$ is in the fundamental
region, i.e. is the absolute minimum of 
$\frac{1}{2} \int d^{D}x \;{\rm Tr} (A_{i}^{g})^{2}$
with respect to $g$. In \rf{sphere}, the
gauge transformation $h$ minimizing the
second integral can be assumed to be of the form $e^{i\delta H}$, where the
Lie-algebra valued field $\delta H$
is infinitesimal. Let us define a new variation of $A$, which we call for the moment
$\delta A^{\prime}$ given by
\beq
A_{j}+\delta A^{\prime}_{j}=(A_{j}+\delta A_{j})^{h}
=A_{j}+\delta A_{j}-i[\delta H, A_{j}+\delta A_{j}]+
i\partial_{j}\delta H+\dots \;.\nonumber
\eeq
In general, the new gauge connection $A+\delta A^{\prime}$ remains in the
fundamental domain of orbit space
(the exceptional situations
occur when $A$ is at the boundary of the fundamental
region). It is
a particular gauge connection lying within the orbit $\beta$. In 
particular, $A+\delta A^{\prime}$ satisfies Coulomb gauge and half the 
integral of its trace squared is equal to $\rho_{0}^{2}$:
\beq
\partial_{j}\delta A_{j}^{\prime}=0\;,\;\;
\rho_{0}^{2}=\frac{1}{2} \int d^{D}x\; {\rm Tr} A_{j}^{2}=\frac{1}{2} 
\int d^{D}x\; {\rm Tr} (A_{j}+\delta A_{j}^{\prime})^{2} \; . 
\label{someconditions}
\eeq
For convenience, we now relabel $\delta A^{\prime}$ by $\delta A$.

We must impose the 
constraints \rf{someconditions} when investigating the variation of the
potential energy $U$. This is 
because we wish to see how variations $\delta A$ subject 
to the condition that the gauge orbit remains on the sphere of radius $\rho_{0}$
change $U$. If such extremal variations produce only positive
changes in $U$, then $\alpha$ lies on a river valley. Our problem has been reduced to
the study of the functional 
\beq
J[A;\;\delta A]
&\equiv& U[\beta]-U[\alpha] =
 \frac{\lambda}{2} \int d^{D}x \;( \delta A_{j}^{a})^{2}  \nonumber \\
&+&\frac{1}{2} \int d^{D} x \;
 \delta A_{j}^{a} \;\left[\; -({\cal D}^{2})^{ab}\delta_{jk}+({\cal D}_{j}
    {\cal D}_{k})^{ab} +2C^{acb}F_{jk}^{c}\;\right]\; \delta A_{k}^{b}\; ,
\label{HamJac}
\eeq
where $\delta A$ is subject to the conditions (equivalent to \rf{someconditions})
\beq
\partial_{j}A_{j}^{a}=0\;,\;\;\int d^{D}x\, \left[ 2(A_{j}^{a}\delta 
A_{j}^{a})^{2} +(\delta A_{j}^{a})^{2}\right]=0 \label{someconditions1}
\eeq
We have used the fact that $A$ satisfies the Proca equation and then used the second of
equations \rf{someconditions1}
to simplify 
the first term on the right-hand side of \rf{HamJac}.  The precise formulation of our question is the following: is $J[A;\;\delta A]$ positive for
variations $\delta A$ satisfying \rf{someconditions1}?


\section{SU(2) in $2+1$ dimensions}

In the remainder of this paper, we will only consider the case of two
space and one time dimension.

A non-Abelian gauge theory with no matter in two space dimensions is
simpler than a theory in three space dimensions. The $2+1$-dimensional
Yang-Mills theory is ultraviolet
finite and therefore the bare dimensionful coupling constant can be fixed
to some finite nonzero value. The metric and the potential energy are not
renormalized by infinite constants. We take gauge group $G={\rm SU(2)}$ and
generators $t^{a}=\sigma^{a}/2$, where $\sigma^{1}$, $\sigma^{2}$ and
$\sigma^{3}$ denote the three Pauli matrices.  The structure coefficients
are therefore $C^{abc}=\epsilon^{abc}$. 

We shall find it necessary to impose an infrared cutoff.  To illustrate
the reason for this, consider the problem of extremizing (\ref{extrem})
when the space is the infinite plane.  Consider an arbitrary connection
$A_i(x)$.  Under the scale transformation, $A_i(x)\rightarrow
A^\prime_i(x)=sA_i(sx)$, where $s$ is a positive real number, the orbit
$\alpha$ containing $A_i(x)$ is mapped to an orbit $\alpha^\prime$ (for
further discussion, see Section 10 of reference \cite{orland}).  Under this
transformation of orbits, $\rho_0$ is unchanged but $U(\alpha)$ changes by
an overall factor $U(\alpha')=s^2 U(\alpha)$.  Therefore, we can transform
the potential energy to a value as small as desired for a given value of
$\rho_0$.  For this reason, Yang-Mills theory on the infinite plane has no
non-trivial extremal curves with finite $\rho_0$. Moreover, to lower its
potential energy and preserve its value of $\rho_0$, a gauge field tends
to spread out to infinite size.  In order to cutoff this infrared
behavior, we will put the system in a box with periodic boundary
conditions, i.e. a torus with coordinates $0\le x^{i} < L_{i}$, for
$i=1,2$ and all functions of $x$ to be doubly-periodic with periods
$L_{1}$ and $L_{2}$.

We do not consider twisted boundary conditions in this paper
\cite{thooft}. However, we should like to mention
that the case of a twist can easily
be incorporated by doubling one of our torus
dimensions $L_{1}$ or $L_{2}$. Any SU(2) Yang-Mills-Proca solution on the twisted
torus of dimensions $L_{1}$ by $L_{2}$ is automatically a Yang-Mills-Proca
solution on the periodic torus of dimensions $L_{1}$ by $2L_{2}$.

\section{Embedded Abelian curves}
\setcounter{equation}{0}
\renewcommand{\theequation}{5.\arabic{equation}}

The first set of solutions to the Proca equation \rf{massiveYM} we will
consider are those which lie entirely in an Abelian subalgebra,
$A_{i}(x)=A_{i}^{1}(x) t^{1}$. The most general solutions of this kind on
the torus are simple to find: 
\beq
A_{1}^{1}(x)=q_{2}\;(\cos q_{1}x^{1}, \sin q_{1}x^{1} )\; {\rm M} \;\left( 
\begin{array}{c} -\sin q_{2}x^{2} \;,\\
\cos q_{2}x^{2} \end{array} \right)   \;, \nonumber \\
A_{2}^{1}(x)=q_{1}\;(\sin q_{1}x^{1}, -\cos q_{1}x^{1} )\; {\rm M} 
\;\left( \begin{array}{c} \cos q_{2}x^{2} \\
\sin q_{2}x^{2} \end{array} \right)  \;,
\nonumber \\
\lambda=-q_1^2-q_2^2\;,~~~~~~~~~~~~ \label{Abelian}
\eeq
where $q_{i}=\frac{2\pi l_{i}}{L_{i}}$, $l_{i}$ are integers, the
case $q_{1}=q_{2}=0$ is excluded
and M is an arbitrary real two-by-two matrix.

It is clear that \rf{Abelian} is precisely of the form of the river
valleys of an Abelian gauge theory. Calculating $\rho_{0}$ and $U$ is
quite easy and reveal a harmonic oscillator potential for each choice
of $q_{i}$. The excitations are simply those of the oscillators; they
are photons. The oscillator frequency is minimized by taking one of 
the $q_{i}$ equal to zero and the other equal to its smallest allowed
value, i.e. $\frac{2\pi l_{i}}{L_{i}}$. The gap between the ground
state and the first excited state vanishes in the thermodynamic
limit. However, for the SU(2) theory, the situation is quite
different.

For the solution in (\ref{Abelian}), the potential energy is
$$
U=\frac{1}{16}(q^2)^2L_1L_2 {\rm Tr}({\rm MM}^T)
$$
and
$${\cal I}(A,1)
=\frac{1}{2}\int d^2 x {\rm Tr}[A_i(x)]^2= \frac{1}{16}L_1L_2 q^2{\rm 
Tr}({\rm MM}^T)\;. $$ In the region
where ${\cal I}(A,1)$ can be identified with
$\rho^2_0$, the potential energy behaves as 
\begin{equation}
U= q^2\rho^2_0  \;,
\label{potris}
\end{equation}
which grows quadratically with $\rho_0$.  This is always the case 
for the gauge group U(1); it 
is easy to see that any gauge copy of $A$ under Abelian gauge 
transformations ($A\rightarrow A+t^1\nabla\chi$) has a 
larger value of the integral: 
$$
{\cal I}(A,\exp(i t_1\chi))\geq {\cal I}(A,1) \;.
$$
However,
in the SU(2) gauge theory, it is necessary to check whether non-Abelian
gauge copies of $A$ can have a smaller value of the integral, i.e. whether
$g(x)$ exists such that ${\cal I}(A,g)\leq {\cal I}(A,1)$.  If this is 
the case, then $\rho^2_0$ is smaller than ${\cal I}(A,1)$  and the  
potential energy rises faster with increasing $\rho_0$ 
than the quadratic behavior in \rf{potris}.

One approach to this problem is to check whether ${\cal I}[A,g]$ is 
a local minimum at $g(x)=1$.  Indeed, since $A_i(x)$ satisfies the 
Coulomb gauge condition, it is guaranteed to be an extremum.  To determine 
whether this extremum is a local minimum, we must examine the
spectrum of the quadratic form in the Faddeev-Popov functional 
\rf{faddeev}. Indeed, Taylor expanding ${\cal I}[A,g]$ in the 
Lie-algebra valued quantity 
$h=h^{a}t^{a}$ where $g=e^{ih}$ yields \cite{gribov2}
\beq
{\cal I}[A,g]={\cal I}[A,1]\;+\; {\rm F.P.}
={\cal I}[A,1]\;+\;\frac{1}{2}\int d^{2} x\; h^{a}(x)
\;({ \cal M} )^{ab}\; h^{b}(x) \;
+\; \dots \;, \nonumber
\eeq
where ${\cal M}$ denotes the Faddeev-Popov operator
\beq
({\cal M} )^{ab}=-\partial^{2} \delta^{ab}-\epsilon^{c a b}A_{i}^{c}(x)
\partial_{i}\;. \nonumber
\eeq
We will show that this operator always has a vanishing eigenvalue for
the connection \rf{Abelian}. This means that this connection lies on the
so-called Gribov horizon.

We first diagonalize ${\cal M}$ in the color indices to obtain
\beq
{\cal M} = \left( \begin{array}{ccc} 
-\partial^{2} &     0           &     0 \\
0             & -\partial^{2} +iA^1(x)\cdot\partial  & 0 \\
0             &     0          &  
-\partial^{2}-iA^1(x)\cdot\partial\end{array} \right)\;, \label{FP1}
\eeq
The following straightforward variational argument shows that 
$-\partial^{2}\pm iA^1(x)\cdot\partial$ will have 
a vanishing eigenvalue for any choice of the matrix M. In other
words, the connections
\rf{Abelian} all sit on the Gribov horizon. Consider the normalizable
trial function
\beq
\Psi= e^{iR_{1}x^{1}+iR_{2}x^{2}}\left[ 1+(\cos q_{1}x^{1},\sin q_{1}x^{1})
{\rm P} \left( \begin{array}{c} \cos q_{2}x^{2} \\ \sin q_{2}x^{2}
\end{array} \right) \right]
\eeq
where P is a two-by-two complex matrix and
$R_{i}=\frac{2\pi s_{i}}{L_{i}}$, $s_{i}$ an integer. Consider
the matrix element of one of the components
of
the operator \rf{FP1}
\beq
{\rm G}_{\pm}&\equiv& \frac{1}{L_1 L_2}\int d^2x \Psi^*(x)[ -\partial^2\pm iA^1(x)
\cdot\partial ] \Psi(x)= R^{2}+\frac{1}{2}(R^{2}+q^{2}) {\rm Tr P}^{\dagger}
{\rm P}  \nonumber \\
&\pm&  \frac{1}{2}{\rm Tr P}^{\dagger} \left[
R_{2}q_{1}\left( \begin{array}{cc} -{\rm M}_{21} &-{\rm M}_{22}\\
{\rm M}_{11} & {\rm M}_{12}
\end{array} \right)
-R_{1}q_{2}\left( \begin{array}{cc} -{\rm M}_{12} &{\rm M}_{11}\\
-{\rm M}_{22} & {\rm M}_{21}
\end{array} \right)
\right]
\nonumber \\
&\pm&\frac{1}{2}{\rm Tr} \left[
R_{2}q_{1}\left( \begin{array}{cc} -{\rm M}_{21} &{\rm M}_{11}\\
-{\rm M}_{22} & {\rm M}_{12}
\end{array} \right)
-R_{1}q_{2}\left( \begin{array}{cc} -{\rm M}_{12} &-{\rm M}_{22}\\
{\rm M}_{11} & {\rm M}_{21}
\end{array} \right)
 \right] {\rm P} \;.  \nonumber
\eeq
It is straightforward to minimize this expression with respect to P
by completing the square. The smallest value of ${\rm G}_{\pm}$ is
\beq
\min {\rm G}_{\pm}=R^{2}+\frac{1}{8(R^{2}+q^{2})}
\left[
(R_{1}^{2} q_{2}^{2}+R_{2}^{2}q_{1}^{2})\; {\rm Tr\,\, M}^{T} {\rm M}
+4R_{1}R_{2}q_{1}q_{2}\; {\rm det\,\, M} \;.
\right]
\eeq
Among our choices of $\Psi$, there is always one for which this expression
is zero, namely that with $R_{1}=R_{2}=0$. While the connection
\rf{Abelian} is a solution of the Proca equation, it lies on the Gribov
horizon for all M. Therefore, these connections do not lie in the fundamental
region and do not constitute an extremal curve.

We have found a striking difference between the Abelian and non-Abelian
gauge theory. In the Abelian theory, standing-wave connections lie in the
orbits of true river valleys. Excitations in these river valleys
are photons. In the Yang-Mills theory, such connections are not
extremal curves. This result strongly suggests that the excitations
are not perturbative gluons.

\section{River valleys containing flat connections}
\setcounter{equation}{0}
\renewcommand{\theequation}{6.\arabic{equation}}

An obvious strategy to find minima of the potential is to look for
configurations of zero field strength. Not all such configurations are
pure gauge \cite{thooft}. The idea is to find solutions of the field
equations $F=0$ which are at a particular distance from the trivial
configuration $A=0$, 
\beq
\rho_0^2=\inf_g \frac{1}{2}\int
d^2x{\rm Tr}\left[A^g(x)\right]^2 \;,\nonumber 
\eeq
where $A^g$ is
a static gauge transform of $A$. The space of gauge orbits containing 
such connections is automatically a river valley, since the potential
energy has saturated its lower bound, namely zero.

Consider solutions of the condition $F=0$ on the torus.   A connection is
flat when it is written as
\begin{equation}
A_i(x)=ig^{-1}(x)\partial_i g(x)\;, \label{flatcon}
\end{equation}
where $g(x)$ is a unitary matrix.
A gauge transformation
$$
A_i(x)\rightarrow ih^{-1}(x)\left(\partial_i-iA_i(x)\right)h(x) \;,
$$
is equivalent to
$$
g(x)\rightarrow g(x)h(x) \;.
$$
These are trivially solutions of the Proca equation \rf{massiveYM}.

The gauge field must satisfy the boundary conditions of the torus,
\begin{eqnarray}
A_i(x^1+L_1,x^2)=A_i(x^1,x^2)\;,\cr
A_i(x^1,x^2+L_2)=A_i(x^1,x^2) \;.\nonumber
\end{eqnarray}
This will occur if $g(x)$ in \rf{flatcon} obeys the
condition
\begin{eqnarray}
g(x^1+L_1,x^2)=u_1 g(x^1,x^2)\;, \cr
g(x^1,x^2+L_2)=u_2 g(x^1,x^2) \;,
\nonumber
\end{eqnarray}
where $u_1$ and $u_2$ are constant unitary matrices.  The consistency
condition
\begin{eqnarray}
g(x^1+L_1,x^2+L_2)&=&u_1g(x^1,x^2+L_2)=u_1u_2g(x^1,x^2)\;,\cr
g(x^1+L_1,x^2+L_2)&=&u_2g(x^1+L_1,x^2)=u_2u_1g(x^1,x^2)\;,
\end{eqnarray}
requires that $u_1$ and $u_2$ commute.

The expression \rf{flatcon} is unchanged if we replace
$g(x)$ by $vg(x)$ where $v$ is a constant unitary matrix.  Under this
replacement $u_1$ and $u_2$ are replaced by $vu_1v^{-1}$ and
$vu_2v^{-1}$, respectively.   Since $u_1$ and $u_2$ commute, they
can be simultaneously diagonalized by a judicious choice of $v$.
Thus, for SU(2) they can be taken to have the form
\begin{equation}
u_i=\left( \matrix{ e^{i\phi_i} & 0\cr 0 & e^{-i\phi_i}\cr}
\right)\;,
\nonumber
\end{equation}
where the phases lie in the fundamental 
region
\begin{equation}
\left|
\phi_i\right|\leq \pi \;.
\nonumber
\end{equation}

A gauge potential which corresponds to a particular $u_i$ is
\beq
A_i(x)=g^{-1}(x)\left( i\partial_i-2\phi_i t^3/L_i\right) g(x)\;,
\eeq
where $g(x)$ satisfies periodic boundary conditions, $g(x^1+L_1,x^2)=
g(x^1,x^2+L_2)=g(x^1,x^2)$.  Furthermore, 
\beq
\rho_0^2=\inf_h\frac{1}{2}\int d^2x {\rm Tr}\left[ (gh)^{-1}
\left( i\partial_i-2\phi_i t^3/L_i\right)(gh)\right]^2 \;.
\eeq
Though $g(x)$ is periodic on the torus, $h(x)$ need not be. The
transformation $h(x)$ can be 
periodic up to an element of the center of SU(2).  For example, we can 
choose $h(x)$ so that
$$
g(x)h(x)=\exp\left( i\sum_{i=1}^2 2\pi n_i x^i t^3/L_i\right)\;.
$$
This yields an upper bound on $\rho_0^2$,
\beq
\rho_0^2\leq \inf_{n_i} L_1L_2\sum_{i=1}^2 \left( \frac{n_i \pi+\phi_i}
{L_i}\right)^2\leq\frac{\pi^2}{4}\left(\frac{L_1}{L_2}+
\frac{L_2}{L_1}\right)\;.
\eeq
The Gribov
horizon is located where the inequality is saturated. Note
that, if we had instead required that the gauge transformations be
strictly periodic
on the torus, the above equation would read $\rho_0^2\leq \pi^2\left(
\frac{L_1}{L_2}+\frac{L_2}{L_1}\right)$.

\section{Constant-magnitude curvature solutions I}
\setcounter{equation}{0}
\renewcommand{\theequation}{7.\arabic{equation}}

In this section we will find solutions of the Yang-Mills-Proca
equations with the supplementary condition that the field strength
has constant magnitude. What is remarkable is that it is possible to
find {\bf all} such solutions. We answer the question of which of these
are extremal curves in the next section.

We will start with the simplest case, where the field strength
is simply constant. Later (in the next section) we shall see that this case 
is an example of the
general solution. The equations \rf{massiveYM} and \rf{constraint} simplify
considerably if $F_{ij}$ is assumed to be a constant. It will turn out
that these are the {\bf only} constant-magnitude field strength solutions
which are extremal curves.

We begin by looking for a solution for which 
\beq
F_{ij}=\epsilon_{ij} f t^{3}\;, \nonumber
\eeq
with $f$ constant. We have two reasons for treating this case first. The 
first
is that it is considerably easier than the more general situation of
curvature of constant magnitude. The second reason is that, as we show
in Section 8, these 
configurations are the only constant-magnitude curvature
extremal curves, and
therefore it seems worthwhile to derive them simply.

The equation \rf{massiveYM} becomes
\beq
\epsilon^{a\,b\,3}\epsilon_{i\,j}f A_{i}^{b}+\lambda A_{j}^{a}=0 \;, 
\label{constF}
\eeq
from which $A_{j}^{3}=0$ immediately follows.

Define $a$ to be the real $2\times2$ matrix whose $b$ - $i$ component 
is $A_{i}^{b}$. Then 
\rf{constF} becomes
\beq
\sigma^{2}\; a\; \sigma^{2}=\frac{\lambda}{f} \;a \;.\label{constF1}
\eeq
There are solutions of \rf{constF1} only if $\lambda=\pm f$. 
If $\lambda=f$, then
\beq
a=a_{+}\equiv a_{0} {\bf 1}+ia_{2}\sigma^{2}\;, \nonumber
\eeq
while if $\lambda=-F$, then
\beq
a=a_{-}\equiv a_{1}\sigma^{1}+a_{3}\sigma^{3}\;. \nonumber
\eeq
These results may be written in terms of the corresponding 
connections $A_{\pm\,i}$
as
\beq
A_{+,1}&=&a_{3}t^{1}+a_{1}t^{2}\;,\;\; A_{+,2}=a_{1}t^{1}-a_{3}t^{2}\;,
\nonumber \\
A_{-,1}&=&a_{0}t^{1}-a_{2}t^{2}\;,\;\; A_{-,2}=a_{2}t^{1}+a_{0}t^{2}\;.
\nonumber 
\eeq

The curvature must given by
\beq
f=-\frac{1}{2} {\rm Tr}\;
 a_{\pm} \sigma^{2} a_{\pm} \sigma^{2} =\mp \frac{1}{2} {\rm Tr}\;
 a_{\pm}^{2}\;, \nonumber
\eeq
and the Abelian piece $\partial_{1} A_{2}-\partial_{2} A_{1}$ must vanish. This latter 
condition means that
for $\pm=+$, 
$a_{+}^{2}\equiv a_{0}^{2}+a_{2}^{2}$ is constant, while if $\pm=-$, $a_{-}\equiv a_{1}^{2}+a_{2}^{2}$ 
is constant. The Coulomb gauge condition \rf{coulomb1}
then implies that
the numbers 
$a_{1}$, $a_{2}$, $a_{3}$ and $a_{4}$ are 
everywhere harmonic functions, $\partial^{2}a_{q}=0$, $q=1,2,3,4$. Since
these functions are doubly-periodic, Liouville's theorem implies that they
are constants.

To summarize the results obtained in this section thus far, we have
found that the solutions to the Proca equation with
constant field strength are of the form
\beq
A_{1}^{a}=f^{1/2} \delta^{a\,1}\;,\;\;
A_{2}^{a}=f^{1/2} \delta^{a\,2}\;,   \label{constsoln}
\eeq
up to global gauge rotations. Such constant non-Abelian potentials
were discussed long ago by Brown and Weisberger \cite{brown}

Recall that $\rho_{0}^{2}$ is the minimum of 
${\cal I}[A,g]=\frac{1}{4}\int d^{2}x\{[(A^{g}_{1})^{a}]^{2}+[(A^{g}_{2})^{a}]^{2}\}$
with respect to gauge transformations $g$. Each of our constrained Proca
solutions is a local extremum of ${\cal I}[A,g]$ at
$g=1$. However, further work needs to be done to check to see whether
such a solution is an
extremal curve
configuration; in other words, that it is the absolute minimum of the integral of the square of the gauge field, so
that
${\cal I}[A,1]=\rho_{0}^{2}$. Given
${\cal I}[A,1]$ it is possible to determine $F$ and $\lambda$. We find
that for both
$f=\lambda$ and
$f=-\lambda$,
\beq
\lambda=-\frac{2{\cal I}[A,1]}{L_{1}L_{2}} \;. \nonumber
\eeq
An extensive analysis to determine whether
${\cal I}[A,1]$ can be decreased by a gauge transformation
will be done in Section 9. The
gauge orbit of only
one of the resulting solutions (up to translations
and rotations) is found to lie on on an extremal curve.

Next we turn to the more general case of solutions of 
\rf{massiveYM} for which the curvature is not assumed constant, but
the magnitude of the curvature is assumed constant. The key to 
finding these solutions is the
analyticity
of certain quadratic polynomials of the connection. Liouville's theorem
then guarantees that these quadratic polynomials are constants. This enables us 
to use a particular parametrization of the connections and the curvature. The constant-curvature
solutions of the last section are a special case of those we find here.

The equations \rf{massiveYM} may be written as
\beq
\pr_{i}F^{a}+\eps^{a\,b\,c}A^{b}_{i}F^{c}=\l \eps_{ij}A^{a}_{j}\;, \;\;
F_{ij}^{a}\equiv \eps_{i\,j} F^{a} \;.  \label{e of m}
\eeq
We supplement these equations with the condition 
\beq
(F^{a})^{2}\equiv f^{2}\;. \label{constr}
\eeq
We call \rf{e of m}
together with \rf{constr} the {\bf constrained 
Proca
equations}.

The general solutions are found in the appendix. These have spatial 
dependence on only $x^{1}$ or $x^{2}$. If we choose the dependence to
be on $x^{1}$, these have the form
\beq
A_{1}^{a}&=&\frac{f}{\vert \lambda\vert^{3/2}} \gamma_{0}^{a} \;, \nonumber\\
A_{2}^{a}&=&\frac{1}{\vert \lambda \vert^{1/2}}\beta_{0}^{a} \cos
\frac{(f^{2}-\lambda^{2})(x^{1}-x^{1}_{0})}{\vert\lambda\vert^{3/2}}
-\frac{1}{\vert \lambda \vert^{1/2}} F_{0}^{a}\sin\frac{(f^{2}-
\lambda^{2})(x^{1}-x^{1}_{0})}{\vert \lambda\vert^{3/2}}
\;,\nonumber \\
F^{a}&=&\beta_{0}^{a} 
\sin\frac{(f^{2}-\lambda^{2})(x^{1}-x^{1}_{0})}{\vert \lambda\vert^{3/2}}
-F_{0}^{a}\cos\frac{(f^{2}-
\lambda^{2})(x^{1}-x^{1}_{0})}{\vert \lambda\vert^{3/2}}\;, 
\label{curve1}
\eeq
where $F^{a}_{0}=F^{a}(x^{0})$, and the isovectors
$\gamma^{a}_{0}$, 
$\beta^{a}_{0}$ satisfy $(\gamma^{a}_{0})^{2}=(\beta^{a}_{0})^{2}=f^{2}$, 
and
$\beta^{a}_{0}=\frac{1}{f}\epsilon^{abc}F^{b}\gamma^{c}$. Notice
that this solution
\rf{curve1} is 
consistent with periodic boundary conditions, provided \beq
\frac{f^{2}-\lambda^{2}}{\vert\lambda\vert^{3/2}}=\frac{2\pi n}{L_{1}}\;, 
\label{extrempbc}
\eeq
where $n$ is an integer. This is a quartic equation
for $\vert \lambda \vert$ in terms of the integer
$n$ and has one real root. Thus the solutions on the
torus are quantized for each $f$.

The field strength $F^{a}$ in \rf{curve1} can always be made equal to 
$f\delta^{a3}$ by a gauge transformation, no
matter what the choice of $\vert \lambda \vert$. However, the 
gauge orbit containing the gauge connection
in \rf{curve1} is different for each $\lambda$. Though
the field strength for two such gauge fields, each with 
a different $\lambda$ is the same, it is in general impossible to gauge
transform one gauge field to the other (this will be shown in 
the next section). This 
is an example of the Wu-Yang
ambiguity \cite{wu-yang}.

We now have all the solutions to the
Proca equations which have constant potential-energy density. We note that
these include the constant-field-strength configurations of
\rf{constsoln}. These are simply \rf{curve1} when
$\vert \lambda \vert=f$, i.e. global gauge rotations of
\rf{constsoln}. We now need to
see under which circumstances these are absolute minima of 
${\cal I}[A,g]=\frac{1}{4}\int d^{2}x\{[(A^{g}_{1})^{a}]^{2}+[(A^{g}_{2})^{a}]^{2}\}$ 
at $g=1$.

\section{Constant-magnitude curvature solutions II}

Any river-valley gauge orbit of constant potential
energy density must contain a gauge configuration satisfying
the constrained Proca equations \rf{e of m} and
\rf{constr}. We proved in the appendix that such a connections
must always be of the
form
\rf{curve1}. In this
section
we determine which of the orbits containing \rf{curve1} are in
river valleys. We begin this analysis by working in the full
plane ${\cal R}^{2}$, where
it is somewhat easier than on the 
torus. {\it In the plane we shall find that none
of the solutions} \rf{curve1}{\it are river valleys}. This
conclusion is not so surprising, for without a careful restriction 
on gauge orbits \cite{orland}  
the sphere of constant $\rho_{0}$ is
not a compact space. We then 
show that on the torus only one of the gauge configurations $A_{i}^{a}$ 
given by \rf{curve1} is a local minimum of 
${\cal I}[A,g]$ at $g=1$. However for sufficiently large $f$, at the
Gribov horizon, even
this configuration is not a local minimum. The horizon is reached
at a finite value of $\rho_{0}$.

\begin{center}
{\bf 8.1 The effects of gauge transformations in the plane}
\end{center}
\setcounter{equation}{0}
\renewcommand{\theequation}{8.1.\arabic{equation}}

We first examine \rf{curve1} in the plane 
${\cal R}^{2}$. We shall find that for any of these connections
there always exists a $g(x)\in {\rm SU(2)}$,
$g(x)\neq 1$ such
that 
$[(A^{g})_{1}^{a}]^{2}+
[(A^{g})_{2}^{a}]^{2}<
(A_{1}^{a})^{2}+(A_{1}^{a})^{2}$. The expression on each side
of this inequality is constant, so this is similar to saying 
that ${\cal I}[A,g]<{\cal I}[A,1]$, though
in truth, neither integral exists. A careful definition of gauge orbits
excludes
gauge fields
which are not
square-integrable (clearly \rf{curve1} is not square-integrable in the 
plane). Nonetheless, it is useful to examine $[(A^{g})_{1}^{a}]^{2}+
[(A^{g})_{2}^{a}]^{2}$
for \rf{curve1} in ${\cal R}^{2}$, as it will shed light on the 
situation on the torus. Since
this quantity happens to be independent of $x$ in this subsection, it is 
analogous to the integral
${\cal I} [A,g]$ in the general case.

We should like to mention that the $f>\vert \lambda\vert$ 
connections are gauge equivalent to the $f<\vert \lambda\vert$ 
connections, with the space-coordinate axes interchanged. To show this, we
make a rigid (i.e. independent of $x$)
gauge transformation of \rf{curve1} so that 
$\gamma_{0}^{a} \rightarrow f\delta^{a1}$, $\beta_{0}^{a} 
\rightarrow f\delta^{a2}$ and
$F_{0}^{a} \rightarrow f\delta^{a3}$ and to make a translation so that 
$x^{1}_{0}=0$. We 
then find for the two-by-two Hermitian
matrices $A_{i}$ and $F$
\beq
A_{1}(x)=\frac{f^{2}}{2\vert \lambda\vert^{3/2}} \left( \begin{array}{cc}
0&1\\
1&0 \end{array} \right)\;
&,&\;\;
   A_{2}(x)=\frac{f}{2\vert \lambda\vert^{1/2}} \left( \begin{array}{cc}
   -\sin kx^{1}& -i\cos kx^{1}\\
   i\cos kx^{1}& \sin kx^{1} \end{array} \right)\;, \nonumber \\
   F(x)
&=&\frac{f}{2} 
   \left( \begin{array}{cc}
   \cos kx^{1}& -i\sin kx^{1}\\
   i\sin kx^{1}& -\cos kx^{1} \end{array} \right)\;, \nonumber
\eeq
where $k=(f^{2}-\lambda^{2})/\vert \lambda \vert^{3/2}$. If we perform the
gauge transformation \rf{gaugetrans}
\beq
g(x)=   \left( \begin{array}{cc}
   \cos kx^{1}/2& i\sin kx^{1}/2\\
   i\sin kx^{1}/2& \cos kx^{1}/2 \end{array} \right)\;, \label{trans}
\eeq
we can bring $A_{i}$ to a constant 
gauge field $(A^{g})_{i}$, for which the isovectors $(A^{g})_{i}^{a}$ are
\beq
(A^{g})_{1}^{a}=\vert \lambda \vert^{1/2} \delta^{a1}\;,\;\;
(A^{g})_{2}^{a}=\frac{f}{\vert \lambda \vert^{1/2}} \delta^{a2}\;,\;\;
(F^{g})^{a}=f\delta^{a3}\;. \label{constantg}
\eeq
Notice that under $\lambda \rightarrow f^{2}/\lambda$, the $1$-component
and $2$-component of the gauge connection \rf{constantg} are 
interchanged. Thus an 
$f>\vert \lambda\vert$ 
solution and an $f<\vert \lambda\vert$ solution are equivalent under a gauge 
transformation.

Under a gauge
transformation
\rf{trans} it is possible to lower the value 
of $[(A^{g})_{1}^{a}]^{2}+
[(A^{g})_{2}^{a}]^{2}$. We will show this by a nonrigorous 
argument. Considering formally the 
second variation of the
integral of this
expression (i.e. ${\cal I}[A,g]$)
with respect to a small 
gauge transformation $g\approx 1-ih-h^{2}/2$. The result 
is again the Faddeev-Popov
functional \rf{faddeev}
\beq
{\rm F.P.}=\int d^{2}x\; \frac{1}{2}\left[ 
(\partial_{i}h^{a})^{2}+\epsilon^{abc}A_{i}^{a}\partial_{i}h^{b}h^{c}\right]
\;.\label{faddeevfunc}
\eeq
Even though ${\cal I}[A,g]$ 
is not well-defined, the functional 
${\rm F.P.}$ exists for appropriately defined gauge 
transformations. This quantity can be negative
for some choice of $h^{a}$. For by
Fourier
transforming and replacing
$\partial_{i}$ by $ip_{i}$, the 
eigenvalues of the quadratic form \rf{faddeevfunc}
are 
$(p_{i})^{2}$, $(p_{i})^{2}\pm 2f^{1/2}{\sqrt{(p_{i})^{2}}}$. One of 
these eigenvalues 
is negative 
for any $p_{i}$
such that $(p_{i})^{2}<4f$. Since the values of $p_{i}$ are continuous, a
negative eigenvalue is always present.

It is actually quite easy to 
show that the gauge connections \rf{curve1} are different for
different $\vert \lambda \vert <f$ 
(we have already showed
that each $\vert \lambda \vert <f$ solution is gauge equivalent to a
unique $\vert \lambda \vert >f$ solution). By making the 
further gauge transformation $g^{\prime}(x)$,
\beq
g^{\prime}(x)=\exp (i\vert \lambda\vert^{1/2}x \;t^{1}) \;,\nonumber
\eeq
on \rf{constantg}, the transformed component $A_{1}^{a}$ is brought 
to zero, while
\beq
A_{2}^{a}=\frac{f}{\vert \lambda \vert^{1/2}} \left[
\cos\left( \frac{\vert \lambda \vert^{1/2}}{2}x^{1}\right) \;\delta^{a2}\;+\;
\sin \left( \frac{\vert \lambda \vert^{1/2}}{2}x^{1} \right) \;
\delta^{a3}\right] \;. \nonumber
\eeq
Though these axial-gauge-fixed connections are
all distinct, there do exist gauge transformations such that
each has the same field strength $F^{a}$,
explicitly illustrating the Wu-Yang ambiguity \cite{wu-yang}.

\begin{center}
{\bf 8.2 The effects of gauge transformations on the torus}
\end{center}
\setcounter{equation}{0}
\renewcommand{\theequation}{8.2.\arabic{equation}}

We show below that
only the $k=2\pi n/L_{1}=0$ solution
of \rf{curve1} can possibly lie in the fundamental
region, when the inequality
\beq
{\sqrt{5}}>\frac{L_{1}}{L_{2}}>\frac{1}{\sqrt{5}}     \label{inequality}
\eeq
is satisfied. It may be that only the
$k=0$ solution lies within the
Gribov horizon for other choices
of $L_{1}/L_{2}$ as well, but we have not yet
proved this.

The Faddeev-Popov operator ${\cal M}$ for  \rf{curve1} is
\beq
{\cal M}^{bc}(x,y)
&=&
   [-\delta^{bc}\partial^{2}-
   \epsilon^{1bc}\frac{2f^{2}}{\vert \lambda\vert^{3/2}} \partial_{1}
     \nonumber \\
&-&\epsilon^{2bc}\frac{2f}{\vert \lambda\vert^{1/2}}\cos(kx^{1}) \partial_{2}
+\epsilon^{3bc}\frac{2f}{\vert \lambda\vert^{1/2}}\sin(kx^{2}) \partial_{3}
]  \delta^{2}(x-y)      \nonumber
\eeq
We wish to determine the eigenvalues of ${\cal M}$. Define the
unitary matrix $S^{ab}$, acting only on color indices:
\beq
S=\left( \begin{array}{ccc}
e^{ikx^{1}} & 0           & 0 \\
0          & 1/{\sqrt 2} & i/{\sqrt 2} \\
0          & i/{\sqrt 2} & 1/{\sqrt 2}   \end{array} \right)  \;.
\nonumber
\eeq
Then $S^{\dagger}{\cal M} S$ has the same eigenvalues as ${\cal M}$. This
new operator has the form
\beq
(S^{\dagger}{\cal M} S)(x,y)=\;\;\;\;\;\;\;\;\;\;\;\;\;\;\;\;\;
\;\;\;\;\;\;\;\;\;\;\;\;\;\;\;\;\;\;\;\;\;\;\;\;\;\;\;\;\;\;\;\;\;\;\;\;
\;\;\;\;
\nonumber
\eeq
\beq
 \left( \begin{array}{ccc}
-\partial^{2}-2ik\partial_{1}+k^{2}&
-i\frac{{\sqrt{2}}f}{\vert \lambda \vert}\partial_{2}&
-\frac{{\sqrt{2}}f}{\vert \lambda \vert^{1/2}}\partial_{2}\\
-i\frac{{\sqrt{2}}f}{\vert \lambda \vert}\partial_{2}&
-\partial^{2}-2i\frac{2f^{2}}{\vert \lambda\vert^{3/2}}\partial_{1}+k^{2}&
0\\
\frac{{\sqrt{2}}f}{\vert \lambda \vert^{1/2}}\partial_{2}&
0& -\partial^{2}+2i\frac{2f^{2}}{\vert \lambda\vert^{3/2}}\partial_{1}+k^{2}
\end{array} \right)\delta^{2}(x-y)  \;,
\nonumber
\eeq
which contains no functions of the coordinates $x$. Thus this operator
has the same eigenvalues as that with $\partial_{1}$ and $\partial_{2}$
replaced with
$p_{1}=\frac{2\pi j_{1}}{L_{1}}$ and
$p_{2}=\frac{2\pi j_{2}}{L_{2}}$, respectively. The secular determinant
of the resulting matrix vanishes at the eigenvalues of ${\cal M}$. For
each choice of $j_{1}$ and $j_{2}$ there are
three eigenvalues $w(j_{1},j_{2};1)$, $w(j_{1},j_{2};2)$
and $w(j_{1},j_{2};3)$ The
eigenvalues $w(j_{1},j_{2};q)$ are determined by
\beq
w(j_{1},j_{2};q)=p^{2}+W(j_{1},j_{2};q)\;,\;\;\;\;\;\;\;\;\;\;\;\;\;\;\;
\;\;\;\;\;\;\;\;\;\;\;\;\;\;\; \nonumber
\eeq
\beq
(W-2kp_{1}-k^{2})\left( W^{2}-
\frac{4f^{4}}{\vert \lambda \vert^{3}}p_{1}^{2} \right)
-\frac{4f^{2}}{\vert \lambda \vert}p_{2}^{2} W=0\;. \label{eigen}
\eeq
We will show that if \rf{inequality} holds, then there are negative
$w(j_{1},j_{2};q)$ for some $j_{1}$, $j_{2}$ and $q$, unless $n=0$.

First consider the solutions of \rf{eigen} with
$j_{2}=0$. These are
\beq
w(j_{1},0;1)&=&(p_{1}+k)^{2}\;,\;\;
w(j_{1},0;2)=\vert p_{1}\vert \left( \vert p_{1} \vert
+ \frac{2f^{2}}{\vert \lambda \vert ^{3/2}} \right) \;, \nonumber \\
w(j_{1},0;3)&=&\vert p_{1}\vert \left( \vert p_{1} \vert
- \frac{2f^{2}}{\vert \lambda \vert ^{3/2}}\right)\;. \nonumber
\eeq
In order for both $w(j_{1},0;2)$ and $w(j_{1},0;3)$ to be nonnegative
we must have 
\beq
\left\vert \frac{2\pi p_{1}}{L_{1}}\right\vert=
\vert p_{1} \vert \ge\frac{2f^{2}}{\vert \lambda \vert^{3/2}}\ge 0\;,
\nonumber
\eeq
where the second inequality follows from the fact that the right
hand side is nonnegative. Using \rf{extrempbc} to eliminate
$f$ this becomes
\beq
\vert j_{1} \vert \ge 2n+ \frac{L_{1}\vert \lambda \vert^{1/2}}{2\pi}\ge 0\;.
\nonumber
\eeq
If $n$ is any positive integer, this inequality must be violated for
some $j_{1}$. Therefore $n\le 0$, and there is a negative eigenvalue
unless
\beq
1\ge 2n+  \frac{L_{1}\vert \lambda \vert^{1/2}}{2\pi}\ge 0\;.  \nonumber
\eeq
Writing $-n$ as $\vert n \vert$ and adding $2\vert n \vert$ to all
three sides gives the necessary conditions for eigenvalues
of the Faddeev-Popov operator to be nonnegative :
\beq
2\vert n \vert +1 \ge  \frac{L_{1}\vert \lambda \vert^{1/2}}{2\pi}\ge
2\vert n \vert \;,
\;\; n\le 0 \;. \label{neccond1}
\eeq

Next let us examine the case $j_{1}=0$. We shall see that this implies
another set of conditions, which together with \rf{inequality} and
\rf{neccond1} implies $n=0$. The eigenvalues of the Faddeev-Popov
operator ${\cal M}$ are now
\beq
w(0,j_{2};1)&=&p_{2}^{2}\;,\;\;
w(0,j_{2};2)\;=\;p_{2}^{2}+\frac{k^{2}}{2}
+{\sqrt{\frac{k^{2}}{4}+\frac{4f^{2}}{\vert \lambda \vert}p_{2}^{2}}}\;,\;\;
\nonumber \\
w(0,j_{2};3)&=&p_{2}^{2}+\frac{k^{2}}{2}
-{\sqrt{\frac{k^{2}}{4}+\frac{4f^{2}}{\vert \lambda \vert}p_{2}^{2}}}\;.
\nonumber
\eeq
If there are no negative eigenvalues $w(0,j_{2};q)$, then for any $j_{2}$
we must have
\beq
\left(p_{2}^{2}+\frac{k^{2}}{2}\right) \ge \frac{k^{4}}{4} +
\frac{4f^{2}}{\vert \lambda \vert}p_{2}^{2} \;.  \nonumber
\eeq
Eliminating $f$ with \rf{extrempbc} and simplifying gives
\beq
4\left(\frac{\vert \lambda \vert^{1/2}L_{1}}{2\pi}\right)^{2}
-4\vert n \vert\frac{\vert \lambda \vert^{1/2}L_{1}}{2\pi}
-n^{2}-\frac{L_{1}^{2}}{L_{2}^{2}}j_{2}^{2} \le 0  \;, \nonumber
\eeq
which is an inequality for a quadratic polynomial in
$\vert \lambda \vert^{1/2}L_{1}/(2\pi)$ with positive coefficient
in the highest order (quadratic) term. Therefore, the inequality
is satisfied if $\vert \lambda \vert^{1/2}L_{1}/(2\pi)$ is between
the roots of the polynomial, i.e.
\beq
\frac{\vert n \vert}{2}+
{\sqrt{ \frac{n^{2}}{2}+
\frac{ L_{1}^{2} }{ 4L_{2}^{2} }   } }
\ge \frac{ \vert \lambda \vert^{1/2} L_{1} }{2\pi} \ge 
\frac{\vert n \vert}{2}-
{\sqrt{ \frac{n^{2}}{2}+
\frac{ L_{1}^{2} }{ 4L_{2}^{2} }   } } \;.
\label{neccond2}
\eeq

To satisfy both \rf{neccond1} and \rf{neccond2} the left-hand side
of \rf{neccond2} must be greater or equal to the right-hand side
of \rf{neccond1}, i.e.
\beq
\frac{3}{2} \vert n \vert -{\sqrt{\frac{n^{2}}{2}
+\frac{L_{1}^{2}}{4L_{2}^{2}}}}  \ge 0     \;,  \nonumber
\eeq
which impossible for $n\neq 0$ if $L_{1}^{2}/L_{2}^{2}<5$. Thus 
if \rf{inequality} is satisfied, ${\cal M}$
has
a negative eigenvalue for a gauge
connection \rf{curve1}, unless the connection is constant. Hence
only the $n=0$
solution can be an extremal curve. Furthermore
none of the connections \rf{curve1} with $L_{1}$ and
$x^{1}$ interchanged with $L_{2}$ and $x^{2}$, respectively can be extremal
curves if $L_{2}^{2}/L_{1}^{2}<5$ unless the connection is constant.

We have proved that the only constant-magnitude-curvature
solutions of the
Yang-Mills-Proca equation which lie within the Gribov horizon
are constant gauge connections, provided \rf{inequality} is
satisfied. We conjecture that examining other choices of $j_{1}$
and $j_{2}$ will eliminate the non-constant connections, even
without imposing \rf{inequality}.

Incidentally, it is quite simple to establish the Wu-Yang ambiguity on
the torus, i.e. 
to prove that many of the
connections \rf{curve1}
are not equivalent under gauge 
transformations, despite their having the same field strength after
gauge transformations. Consider 
the Wilson loop on a closed line of 
length $L_{1}$, parallel to the $x^{1}$-axis:
\beq
{\rm Tr}\;{\cal P} \exp\; i\int_{0}^{L_{1}} A_{1}(x^{1},x^{2})\,
dx^{1}=2\cos \frac{f^{2}L_{1}}{2\vert \lambda \vert^{3/2}}\;, 
\label{Wu-Yang1}
\eeq
and the Wilson loop on a closed line of length $L_{2}$, parallel to the $x^{2}$-axis:
\beq
{\rm Tr}\;{\cal P} \exp\; i\int_{0}^{L_{2}} A_{2}(x^{1},x^{2})\,
dx^{2}=2\cos \frac{fL_{2}}{2\vert \lambda \vert^{1/2}}\;.
\label{Wu-Yang2}
\eeq
These gauge-invariant quantities clearly depend on $\lambda$. This 
does not prove that {\it all} the 
configurations \rf{curve1} are gauge-inequivalent, because of the 
periodicity of the right-hand sides of
\rf{Wu-Yang1} and \rf{Wu-Yang2}. However, it
is clear that
most of the connections 
\rf{curve1} really are gauge-inequivalent, because the values of 
$\lambda$ are quantized according to \rf{extrempbc}. 

\begin{center}
{\bf 8.3 A river valley of the gauge theory on the torus}
\end{center}
\setcounter{equation}{0}
\renewcommand{\theequation}{8.3.\arabic{equation}}

We have shown that the only possible river-valley gauge connections 
on the torus with 
constant $(F^{a})^{2}$ are given by 
with $\vert \lambda
\vert=f$. These have have, of 
course, constant gauge connections $A_{i}^{a}$. We will show that
these are indeed river-valley configurations 
for $f$ sufficiently small.

Let us first check whether ${\cal I}[A,1]$ 
is a minimum of ${\cal I}[A,g]$ for this
configuration. If this is the case, then 
the distance of the gauge orbit containing $A_{i}^{a}$ from
the pure-gauge orbit is given by 
$\rho_{0}={\cal I}[A,1]$. Let us begin by repeating the arguments
in Subsection 8.1. The second variation
of 
this integral with respect to small gauge transformations
$g\approx 1-ih-h^{2}/2$ is again the Faddeev-Popov
functional \rf{faddeev}. However, the 
integral is now performed over the torus instead 
of the plane. As done for \rf{faddeevfunc}, Fourier
transforming and replacing
$\partial_{i}$ by $ip_{i}$, in \rf{faddeev}, the 
eigenvalues of the 
quadratic form \rf{faddeevfunc}
are $(p_{i})^{2}$, $(p_{i})^{2}\pm 2f^{1/2}{\sqrt{(p_{i})^{2}}}$. The 
new feature is that
$p_{i}$ is quantized, and may only 
take the values $p_{i}=2\pi j_{i}/L_{i}$, for some
integers $j_{1}$ and 
$j_{2}$. With the 
exception of the zero modes of global gauge transformations when
$p_{1}=p_{2}=0$, the eigenvalues
are all positive, if and
only if
\beq
f<\pi^{2} {\rm min}\left\{ \frac{1}{\;L_{1}^{2}}\;, \;\frac{1}{L_{2}^{2}}\;
\right\} \;. \label{positivity}
\eeq 
This inequality implies that the length of this curve is bounded in
the thermodynamic limit:
\beq
\rho_{0}^{2}\le I[A,1]=\frac{fL_{1}L_{2}}{2} <
\frac{\pi^{2}}{2}
\min \left\{\frac{L_{2}}{L_{1}}, \frac{L_{1}}{L_{2}}
\right\}
\;. \label{bound}
\eeq

We will argue, using the methods of Subsection 3.2, that the family of 
gauge orbits \rf{curve1} with $\vert \lambda \vert =f$ is indeed a 
river valley. Let us examine of the 
functional $J[A;\;\delta A]$ defined in \rf{HamJac}. As discussed
in Subsection 3.2, we wish to determine whether 
this functional is positive, subject to the conditions
\rf{someconditions1}. Since the 
gauge connection $A$ is translation invariant, we must actually
uncover
choices of $\delta A$ for 
which $J[A;\; \delta A]$ vanishes. However, under the variations
$\delta A$ orthogonal to these translations, the functional $J[A;\;\delta 
A]$ is positive within the Gribov horizon. This establishes that
the family of gauge orbits parametrized by $f$ is indeed a river valley.

The gauge connections we are considering are of the form
$A_{j}^{a}=f^{1/2}\delta_{j}^{a}$. We
can solve the first of \rf{someconditions1} by writing the variation of 
gauge connection as a Fourier series \beq
\delta A_{l}^{a}(x)= 
\frac{i\epsilon_{lm}}{L_{1}L_{2}}\sum_{p_{i}= {2\pi j_{i}}/{L_{i}}}
p_{m}\,{\cal E}^{a}(p)\,e^{ip\cdot x}\; , \label{fourier}
\eeq
with the choice of wavenumbers $p_{1}=p_{2}=0$ automatically
excluded. Substituting
\rf{fourier} into \rf{HamJac} yields
\beq
J[A;\; \delta A]
&=& 
         \frac{1}{2L_{1}L_{2}}\sum_{p_{i}}\;
        ({\cal E}^{1}(p),{\cal E}^{2}(p),{\cal E}^{3}(p)) \nonumber \\
&\times&\left( \begin{array}{ccc}
(p^{2})^{2}        &  -fp_{1}p_{2}       & -2if^{1/2}p_{2}p^{2} \\
 -fp_{1}p_{2}      & (p^{2})^{2}         & 2if^{1/2}p_{1}p^{2} \\
2if^{1/2}p_{2}p^{2}& 2if^{1/2}p_{1}p^{2} &  (p^{2})^{2}  
\end{array} \right) \left( \begin{array}{ccc}{\cal E}^{1}(p)\\
{\cal E}^{2}(p)\\{\cal E}^{3}(p)
\end{array} \right)\;. \label{quadform}
\eeq
The matrix in \rf{quadform} has eigenvalues
\beq
\omega_{0}=(p^{2})^{2}, \;\omega_{\pm}=(p^{2})^{2}\pm {\sqrt{4f(p^{2})^{3}+
f^{2}p_{1}^{2}p_{2}^{2}  }}\; , \nonumber
\eeq
of which only $\omega_{-}$ can possibly be zero or negative. The
eigenvalue $\omega_{-}$ is positive provided \rf{positivity}
holds.

We have proved that the connections
\rf{curve1} with $\vert \lambda \vert =f$ lie within
the Gribov horizon and are stable under those
variations which do not change $I[A,1]$. What remains to
to show is that they lie within the fundamental region. To prove this
rigorously we would have to establish that there is no choice of
$g$, such that $I[A,g]<I[A,1]$. Though we have not done this, we have not
succeeded in reducing $I$ by a gauge transformation and have convinced
ourselves that this is not possible.  Consequently, we assert that these
gauge connections lie in
gauge orbits $\alpha$, the set of which
is a river valley. From \rf{bound}
we conclude that the $\rho_{0}^{2}$ has
values on this river valley over the interval from zero to
$\frac{\pi^{2}}{2}
\min \left\{\frac{L_{2}}{L_{1}}, \frac{L_{1}}{L_{2}}
\right\}$.

\section{Feynman's arguments and vortex solutions}

We now attempt to compare the results of this
paper to that of Feynman \cite{feynman}. In our
language, Feynman
argued that a slowly-varying connection $A$ in an orbit $\alpha$
has a value of $\rho_{0}$
which is considerably smaller than $I[A,1]$. He thereby concluded that the
lowest-lying excitations were not gluons. This conclusion is strongly
supported
by our results. Feynman also argued that in
reducing $I[A,1]$ to $I[A,g]$, slowly-varying
connections $A$ would be gauge transformed to connections $A^{g}$
with a periodic
structure, of period $\simeq 1/{\sqrt{F}}$. He went on to
argue that the genuine excitations
are described by small oscillating domains of magnetic flux whose
size is equal to this period. However, we
find no evidence of this. We believe the reason
Feynman came to this conclusion is that his method of
lowering the integral $I[A,1]$ to $I[A,g]$ was heuristic and did
not provide a lower bound. He carried out this reduction 
in the thermodynamic limit, starting with connections
of constant field strength of the form
$A_{1}=0$, $A_{2}^{a}=f\delta^{a3}x^{1}$. His gauge-transformed
gauge potential is
not included among the general solutions we have found (in a finite
volume, Feynman's connection is inequivalent to any we consider)
\cite{brown, wu-yang}. We
have found the
smallest $I[A,g]$ for configurations of constant field strength
in the previous two sections and
do not see evidence of a periodic structure in $A^{g}$.

An excitation of the form suggested by Feynman would occur if there
existed a river valley whose elements were described by elliptic solutions
of \rf{massiveYM}, i.e. solutions with $\lambda>0$. In fact
there are such
elliptic solutions; they are non-Abelian vortices
\cite{nielsen}. This is
seen by introducing a chiral field $g$ and writing
\rf{massiveYM}
and \rf{coulomb1} as
$$-[D^{g}_{i},F^{g}_{ij}] + \lambda (A^{g})_{j}=0 \;,$$
$$\partial_{j} (A^{g})_{j} = 0  \;,$$ 
where $A^{g}$, is defined by \rf{gaugetrans} as
before, $D^{g}_{i}=g^{-1}D_{i}g$ and
$F^{g}_{ij}=g^{-1}F_{ij}g$ (We are
reinterpreting our Proca equation as the unitary gauge
formulation of the gauged chiral sigma model). However these
vortex solutions have logarithmically-divergent potential energy. This
energy can be regularized with a cut-off, but in $2+1$ dimensions, this
field theory requires
only finite renormalization, as the
cut-off is removed. Upon removal
of the cut-off, vortices are suppressed.

\section{Conclusions}

In our studies of the structure of the potential-energy functional
on orbit space we have found significant differences between the
SU(2) and Abelian gauge theories. The fact that 
standing-wave Proca solutions of
Section 5 lie on the Gribov horizon is evidence that the
quantized particle
excitations bear no resemblance to perturbative gluons. The
families of gauge orbits of minimal potential energy lie in
river valleys of finite length. If this is true for
all such families of orbits, a mass gap must be present.

We are attempting to construct most of the river valleys for
the $2+1$-dimensional SU(2) theory, at least
approximately. This may be sufficient to find the vacuum and the lowest
excited states. 

We are also studying the extremal-curve problem in $3+1$ dimensions. There
are nontrivial solutions of the Proca equation for the
hyperbolic ($\lambda>0$), parabolic ($\lambda=0$) and
elliptic ($\lambda<0$) cases. Unlike the case of $2+1$ dimensions, the
elliptic solutions (which have ultraviolet-divergent
potential energy) cannot be dismissed out of hand, since the
coupling constant has an infinite renormalization. These elliptic
solutions include vortices \cite{nielsen}, as well as other
configurations.

\section*{Acknowledgements}

We thank C.~K. Zachos for discussions and
V.~P. Nair for drawing our attention to reference \cite{brown}.

P.O. thanks the staffs of the Niels Bohr Institute, Argonne
National Laboratory and the RIKEN/BNL Research Center for their hospitality
and financial support.

\section*{Appendix: the general solution of the constrained Proca equation}

\begin{center}
{\bf A.1 Analyticity and parametrization of connections}
\end{center}
\setcounter{equation}{0}
\renewcommand{\theequation}{A.1.\arabic{equation}}

By virtue of \rf{e of m} and \rf{constr} we have that
\beq
A_{i}^{a} F^{a}=0\; . \label{orthog}
\eeq
From the expression for $F$ in terms of $A$ we see that
\beq
A^{a}_{i}(\pr_{1}A^{a}_{2}-\pr_{2}A^{a}_{1})=0 \;. \nonumber
\eeq
Let us write these equations in complex coordinates, 
$z=x^{1}+ix^{2}$ and ${\bar z}=x^{1}-ix^{2}$, 
$\pr=\pr /\pr z=(\pr_{1}-i\pr_{2})/2$, ${\bar \pr}=\pr /\pr {\bar z}
=(\pr_{1}+i\pr_{2})/2$. The Hermitian gauge connection
becomes non-Hermitian in
these coordinates: 
$A=(A_{1}-iA_{2})/2$, ${\bar A}=(A_{1}+iA_{2})/2$. Then 
\beq
A^{a}(\pr {\bar A}^{a}-{\bar \pr} A^{a})=
{\bar A}^{a}(\pr {\bar A}^{a}-{\bar \pr} A^{a})
=0 \;. \label{curl}
\eeq
The Coulomb gauge condition is
\beq
\pr {\bar A}^{a}+{\bar \pr} A^{a}=0 \; \label{cou}
\eeq
Equations \rf{curl} and \rf{cou} imply
\beq
{\bar \pr} (A^{a})^{2}=
{\pr}({\bar A}^{a})^{2}=0\;, \label{analytic}
\eeq

The meaning of equations
\rf{analytic} is that 
$(A^{a})^{2}$ is a 
complex-analytic function. If doubly-periodic boundary conditions
are imposed, this function must be a constant by Liouville's
theorem. However, it is interesting to the general solution on the
plane. We will see that even without doubly-periodic boundary conditions, $(A^{a})^{2}$
is a constant.

We can derive the 
relations \rf{analytic} another way. If we view our system 
as a Euclidean field theory in two dimensions, the
two-dimensional
energy-momentum tensor 
associated
with \rf{e of m} has components:
\beq
T_{11}&=&\frac{1}{2}(F^{a})^{2}+
         \frac{\lambda}{2}(A_{1}^{a})^{2}-\frac{\lambda}{2}
          (A_{2}^{a})^{2} \;, \nonumber\\
T_{22}&=&\frac{1}{2}(F^{a})^{2}-\frac{\lambda}{2}(A_{1}^{a})^{2}
          +\frac{\lambda}{2}(A_{2}^{a})^{2} \;, \nonumber\\
T_{12}&=&T_{21}=\lambda A_{1}^{a}A_{2}^{a} \;. \nonumber
\eeq
Since $(F^{a})^{2}$ is constant, conservation of energy 
and momentum, $\partial_{i} T_{ij}=0$ implies \rf{analytic}.

The analyticity conditions \rf{analytic} imply that the Yang-Mills
connection has the functional form
\beq
A^{a}(z,\zbar)    &=&Q(z)G_{+}^{a}(z,\zbar)\;, \nonumber \\
\Abar^{a}(z,\zbar)&=&\Qbar(\zbar)
[\cosh\Phi(z,\zbar)G_{+}^{a}(z,\zbar)
+i\sinh\Phi(z,\zbar)G_{-}^{a}(z,\zbar)]
\;, \label{funcform}
\eeq
where $G_{+}^{a}$ is a complex-valued isovector satisfying
\beq
(G_{+}^{a})^{2}=f^{2}\;,\;\; G_{+}^{a}F^{a}=0\;, \label{gplus}
\eeq
and $G_{-}^{a}$ is the complex-valued isovector defined by
\beq
G_{-}^{a}=\frac{1}{f}\epsilon^{abc}F^{b}G_{+}^{c}\;. \label{gminusdef}
\eeq
Notice that $G_{-}^{a}$ automatically satisfies
\beq
(G_{-}^{a})^{2}=f^{2}\;,\;\; G_{-}^{a}F^{a}=0\;. \label{gminus}
\eeq
The isovectors 
$G_{+}^{a}$, $G_{-}^{a}$ and $F^{a}$ are not the basis of a right-handed coordinate
system of ${\bf R}^{3}$, since $G_{+}^{a}$ and $G_{-}^{a}$ are not orthogonal 
(no complex conjugates are present in \rf{gplus}, \rf{gminus})
or even
real. In fact the matrix ${\bf G}$ defined by 
\beq
{\bf G}= \left( \begin{array}{cc} G^{T}_{+} \\
                                  G^{T}_{-} \\
                                  F^{T}   \end{array} \right) 
=\left( \begin{array}{ccc} 
G^{1}_{+} & G^{2}_{+} & G^{3}_{+} \\
G^{1}_{-} & G^{3}_{-} & G^{3}_{-} \\
  F^{1}   &  F^{2}    & F^{3}  \end{array}
    \right)\;, 
\eeq
whose determinant equals unity by \rf{gminusdef}, is not in SO(3) but 
rather in the adjoint representation of SL(2,{\bf C})
(since $F^{a}$ is real, $\bf G$ cannot be an arbitrary 
element of adj[SL(2,{\bf C})]). In any case, the
``vector products" in \rf{gplus} and \rf{gminus} are not inner products. 

The function $\Phi(z,\zbar)$ is not arbitrary, but is chosen so that
\beq
{\bar G}^{a}_{+}(z,\zbar)=\cosh\Phi(z,\zbar)G_{+}^{a}(z,\zbar)
+i\sinh\Phi(z,\zbar)G_{-}^{a}(z,\zbar)  \;. \nonumber
\eeq
Such a choice of $\Phi$ can always be made. One may always write $G_{\pm}^{a}$
as
\beq
G_{+}^{a}(z,\zbar)&=& \gamma^{a}(z,\zbar) \cosh\frac{\Phi(z,\zbar)}{2}
                        -i\beta^{a}(z,\zbar)\sinh\frac{\Phi(z,\zbar)}{2} \;,
                         \nonumber \\
G_{-}^{a} (z,\zbar)&=&\beta^{a}(z,\zbar) \cosh\frac{\Phi(z,\zbar)}{2}
                     +i\gamma^{a}(z,\zbar)\sinh\frac{\Phi(z,\zbar)}{2} \;,
                     \label{realparam}
\eeq
where the isovectors $\beta^{a}$, $\gamma^{a}$ and $F^{a}$ are real orthogonal and normalized, i.e.
$(\beta^{a})^{2}=(\gamma^{a})^{2}=f^{2}$ and $\beta^{a}=\epsilon^{abc} F^{b}\gamma^{c}$. We will
find this expression \rf{realparam} useful when we write down the 
explicit form of the
gauge connection and the field strength.

\begin{center}
{\bf A.2 Parallel transport}
\end{center}
\setcounter{equation}{0}
\renewcommand{\theequation}{A.2.\arabic{equation}}

Before proceeding further, it is convenient to define the
complex one-forms $U$, $V$ and $W$ which describe how the
vectors $G_{\pm}^{a}$ and $F^{a}$ change under translations:
\beq
       \partial G_{+}^{a}&=&U\;\;G_{-}^{a}+V\;F^{a}\;, \;\partial G_{-}^{a}=-U\;G_{+}^{a}+W\;F^{a}\;, 
                            \partial F^{a}=-V\;G_{+}^{a}-W\;G_{-}^{a}\;,\;\; \nonumber\\         
{\bar \partial} G_{+}^{a}&=&\Ubar\;\;G_{-}^{a}+\Vbar\;F^{a}\;, \;\;
{\bar \partial} G_{-}^{a}=-\Ubar\;G_{+}^{a}+\Wbar\;F^{a}\;, \;\;
{\bar \partial} F^{a}=-\Vbar\;G_{+}^{a}-\Wbar\;G_{-}^{a}. \nonumber \\
  \label{differ}
\eeq
The form of \rf{differ} is dictated by \rf{constr}, \rf{gplus} and \rf{gminus}. These one-forms
define a connection in the adjoint representation of $SL(2,{\bf C})$, namely
\beq
{\cal B} =d{\bf G} {\bf G}^{-1}=\left( \begin{array}{ccc} 
0  & U & V \\
-U &0 & W \\
-V &-W& 0   \end{array}
    \right)  \;, \;\;
{\bar {\cal B}} =d{\bar {\bf G}} {\bar {\bf G}}^{-1}=
	\left( \begin{array}{ccc} 
0  & \Ubar & \Vbar \\
-\Ubar &0 & \Wbar \\
-\Vbar &-\Wbar& 0   \end{array}
    \right)  
\;.  \nonumber
\eeq
This connection is obviously 
flat from the definition, i.e.
\beq
[\partial - {\cal B}, {\bar \partial}- {\bar {\cal B}}]=0 \;.
\label{flatness}
\eeq
For later convenience, we define the antisymmetric matrices $l_{1}$, $l_{2}$ and
$l_{3}$ by
\beq
l_{1}=\left( \begin{array}{ccc} 
0  & 0 & 0 \\
0  & 0 & 1 \\
0  &-1 &0   \end{array}
    \right)\;, \;\;
l_{2}=\left( \begin{array}{ccc} 
0 & 0 & -1 \\
0 & 0 & 0 \\
1 & 0 & 0   \end{array}
    \right)\;,\;\;  
l_{3}=\left( \begin{array}{ccc} 
0 & 1 & 0 \\
-1& 0 & 0 \\
0 & 0 & 0   \end{array}
    \right)  \;, \nonumber
\eeq  
so that 
${\B}=Wl_{1}-Vl_{2}+Ul_{3}$ and
${\Bbar}={\Wbar} l_{1}-{\Vbar} l_{2}+{\Ubar} l_{3}$.

If we know ${\bf G}$
at one point $z_{0}$, $\zbar_{0}$ 
and we also
know the connection 
${\cal B}$, ${\bar{\cal B}}$ 
everywhere in the plane, we can determine 
$F^{a}$, $A^{a}$ and $\Abar^{a}$ everywhere. This follows from the formula
for parallel transport of the matrix $\bf G$:
\beq
{\bf G}(z,\zbar)=
{\cal P}e^{\int_{z_{0}}^{z}dz^{\prime}{\cal B}(z^{\prime}, \zbar)}
{\cal P}e^{\int_{\zbar_{0}}^{\zbar}d\zbar^{\prime}{\bar {\cal B}}(z_{0}, \zbar^{\prime})} {\bf G}(z_{0},\zbar_{0})
\label{parallel}
\eeq
The path of the integration in \rf{parallel} can be deformed to any path from $z_{0}$, $\zbar_{0}$ 
to $z$, $\zbar$ since ${\cal B}$, ${\bar{\cal B}}$ is flat.

\begin{center}
{\bf A.3 Imposing the equations of motion}
\end{center}
\setcounter{equation}{0}
\renewcommand{\theequation}{A.3.\arabic{equation}}

Simplifying the equations of motion \rf{e of m} with \rf{differ} yields
\beq
-VG_{+}^{a}-WG_{-}^{a}+\epsilon^{abc}QG_{+}^{b}F^{c}=i\lambda Q G_{+}^{a}\;, \nonumber
\eeq
\beq
-\Vbar G_{+}^{a}-\Wbar G_{-}^{a}&+&\epsilon^{abc}\Qbar (\cosh \Phi G_{+}^{b}+i\sinh \Phi G_{-}^{b})F^{c} \nonumber \\
                                &=&-i\lambda \Qbar (\cosh \phi G_{+}^{a}
+i\sinh \phi G_{-}^{a})
\;. \nonumber
\eeq
From the linear independence of $G_{\pm}^{a}$, $F^{a}$, the equations of motion reduce to
\beq
V=-i\lambda Q\;,\;\; W=-fQ\;, \nonumber
\eeq
\beq \Vbar=i\Qbar(f\sinh \Phi+\lambda \cosh \Phi)\;,\;\;
\Wbar=-\Qbar (\lambda\sinh \Phi+f \cosh \Phi)\;. \label{simp e of m}
\eeq
Notice that the equations of motion alone impose no restriction on
$U$ or $\Ubar$.

\begin{center}
{\bf A.4 Consistency of $F$ as curvature}
\end{center}
\setcounter{equation}{0}
\renewcommand{\theequation}{A.4.\arabic{equation}}

We next impose \rf{curvature} and \rf{orthog} on \rf{funcform}. Since $G_{\pm}^{a}$ and $F^{a}$
are linearly independent, \rf{curvature}
is equivalent to
\beq
0&=&G_{+}^{a}({\bar \partial} A^{a}-
\partial \Abar^{a}+\eps^{abc}\Abar^{b}A^{c}) \;, \nonumber \\
0&=&G_{-}^{a}({\bar \partial} A^{a}-
\partial \Abar^{a}+\eps^{abc}\Abar^{b}A^{c}) \;, \nonumber \\
f^{2}&=&2i\;F^{a}({\bar \partial}A^{a}-
\partial \Abar^{a}+\eps^{abc}\Abar^{b}A^{c}) \;. \nonumber
\eeq
Using \rf{orthog} and \rf{constr}, these equations simplify to
\beq
0&=&\Qbar \sinh\Phi\; (i\partial \Phi+U) \;, \label{curvcons1} \\
0&=&Q\Ubar-\Qbar U \cosh\Phi \;(i\partial \Phi+U) \;, \label{curvcons2} \\
-\frac{i}{2}&=&Q\Vbar-\Qbar (\cosh\Phi \;V+ i\sinh\Phi \;W)- iQ\Qbar f\sinh\Phi \;. \label{curvcons3}
\eeq

Substituting the equations of motion \rf{simp e of m} into \rf{curvcons1}
and \rf{curvcons2}
means that either
\begin{itemize}
\item I. $\sinh\Phi=0$, $Q\Ubar=\Qbar U$, or
\item II. $i\partial \Phi +U=0$ and $\Ubar=0$.
\end{itemize}
We will see in the next subsection that I. can be eliminated. The third equation \rf{curvcons3} upon
substitution of \rf{simp e of m} becomes
\beq
\frac{1}{2Q\Qbar}=-2\lambda \cosh \Phi-f \sinh \Phi\;. \label{qqbar}
\eeq
Notice that this equation implies that $\Phi$ is real.

By substituting the equations of motion into the expression for the curvature as we have done, the full content of the original
constrained Proca equations
\rf{e of m}, \rf{constr} 
is in \rf{flatness} and \rf{qqbar} together with either I. or II.

\newpage

\begin{center}
{\bf A.5 Flatness and the solution for $\B$ and $\Bbar$}
\end{center}
\setcounter{equation}{0}
\renewcommand{\theequation}{A.5.\arabic{equation}}

We next evaluate the commutator in \rf{flatness} for I. and II. 

If we assume I., then there must exist a function $c(z,\zbar)$ such that
$U(z,\zbar)=Q(z)c(z,\zbar)$ and 
$\Ubar (z,\zbar)=\Qbar (z)c(z,\zbar)$. The flatness condition \rf{flatness} is
\beq
(Q{\bar \partial} c-\Qbar \partial c)\;l_{3}+2i\lambda Q\Qbar (f\;l_{3}+c\;l_{1})=0\;. \nonumber
\eeq
This equation implies that $c=0$ and $2i\lambda f=0$, and is therefore not viable. 

We see that II. must be satisfied. Evaluating \rf{flatness} yields
\beq
\partial {\bar {\partial}} \Phi+Q\Qbar[2\lambda f \cosh\Phi+
(f^{2}+\lambda^{2})\sinh \Phi]=0\;, \label{lax}
\eeq
and is just the Lax pair of this equation. Defining 
\beq
\xi={\sqrt{f^{2}+\lambda^{2}}}\int^{z}Q(z)dz\;,\;\;
{\bar \xi}={\sqrt{f^{2}+\lambda^{2}}}\int^{\zbar}Q(\zbar)d\zbar
\eeq \nonumber
and the new field
\beq
\Psi=\Phi+\tanh^{-1}\frac{2\lambda f}{f^{2}+\lambda^{2}}\;, \nonumber
\eeq
reveals that \rf{lax} is the sinh-Gordon equation
\beq
\frac{\partial^{2} \Psi(\xi, {\bar \xi})}{\partial \xi \partial {\bar \xi}}+
\sinh \Psi(\xi, {\bar \xi})=0 \;. \label{s-G}
\eeq
However, this equation together with condition \rf{qqbar} actually implies that $\Psi$ is a constant. 

The
solution of \rf{qqbar} can be written in terms of $\Psi$ as
\beq
\Psi=\tanh^{-1}\frac{2\lambda f}{f^{2}+\lambda^{2}}-\tanh^{-1}\frac{2\lambda}{f}
-\sinh^{-1}\frac{1}{2Q\Qbar{\sqrt{f^{2}-4\lambda^{2}}}}\;.
\nonumber
\eeq
Substitution into \rf{s-G} yields
\beq
\frac{\partial Q^{3}}{\partial\xi}\frac{\partial \Qbar^{3}}{\partial{\bar \xi}}
=C_{1}[4(f^{2}-4\lambda^{2})Q^{2}\Qbar^{2}-1]^{3/2}
+C_{2}[4(f^{2}-4\lambda^{2})Q^{2}\Qbar^{2}-1]^{2}\;, \nonumber
\eeq
where $C_{1}$ and $C_{2}$ are constants depending only on $f$ and $\lambda$. In order to have nonconstant
solutions, the right-hand side of this equation would have to factorize into a function of $Q$ times
a function of $\Qbar$. However such a factorization is impossible. Therefore $Q$, $\Qbar$ and 
$\Phi$ are all constants. Of course, we would also have found this result by simply 
imposing
doubly-periodic boundary conditions on the analytic function
$Q(z)$ and the antianalytic function $\Qbar(\zbar)$ (by Liouville's
theorem). It is interesting that $Q$, $\Qbar$ and $\Phi$ are constant even without taking boundary conditions
into account.

Our final expressions for $Q$, $\Qbar$ and $\Phi$ are
\beq
Q={\sqrt{ \left\vert \frac{f^{2}-\lambda^{2}}{4\lambda^{3}} \right\vert }}\; e^{i\theta} \;,\;\;
\Qbar=
{\sqrt{ \left\vert \frac{f^{2}-\lambda^{2}}{4\lambda^{3}} \right\vert }} 
\; e^{-i\theta} \;,\nonumber
\eeq
\beq
\Phi=\tanh^{-1}\frac{2\lambda f}{f^{2}+\lambda^{2}} \;. \nonumber
\eeq
From these expressions, we can evaluate the integral of 
$(A_{i}^{a})^{2}$ to find
\beq
{\cal I}[A,1]=-L_{1}L_{2}\frac{f^{4}(f^{2}+\lambda^{2})}{4\lambda^{3}}\;, 
\nonumber
\eeq
from which we see that $\lambda$ must be negative. The connections
$\B$ and $\Bbar$ are
\beq
       {\cal B} &=&   {\sqrt{ \left\vert \frac{f^{2}-
                    \lambda^{2}}{4\lambda} \right\vert }}\; e^{i\theta}\;
                      \left( \begin{array}{ccc} 
                          0        & 0    & -i\lambda \\
                          0        & 0    & -f \\
                          i\lambda & f    & 0   \end{array}
                    \right)  \;,  \nonumber\\
{\bar {\cal B}} &=&{\sqrt{ \left\vert \frac{f^{2}-\lambda^{2}}{4\lambda} \right\vert }} \; e^{-i\theta}
\left( \begin{array}{ccc} 
0        & 0 & -i\lambda \\
0        & 0 & -f         \\
i\lambda & f & 0  \end{array} \right)  \;{\rm sgn}(f^{2}-\lambda^{2}) 
\;.  \label{bexpr}
\eeq

\begin{center}
{\bf A.6 Periodicity of the gauge fields}
\end{center}
\setcounter{equation}{0}
\renewcommand{\theequation}{A.6.\arabic{equation}}

In order to determine the form of the 
gauge fields at all points on the plane, using \rf{parallel}, we must
diagonalize the matrices \rf{bexpr}. These constant 
matrices are simultaneously diagonalizable, as they must be in
order for \rf{flatness} to hold. The diagonalization is
\beq
\B= S \B_{\rm diag} S^{-1} \;, \;\; \Bbar\;=\; S \Bbar_{\rm diag} S^{-1} \;, \nonumber
\eeq
with
\beq
S =\left( \begin{array}{ccc} 
      -f       & -i\lambda                 & -i \lambda \\
      i\lambda &      -f                   & -f \\
      0        & i\sqrt{f^{2}-\lambda^{2}} &-i\sqrt{f^{2}-\lambda^{2}}
      \end{array}    \right) \;. \nonumber
\eeq
Using this diagonalization, \rf{parallel} may be evaluated:
\beq
{\bf G}(z,\zbar)
&=& 
        \left( \begin{array}{ccc} 
    f^{2}-\lambda^{2}\;{\rm c}& 
    i\vert \lambda \vert \;(1-{\rm c})&
    i\vert \lambda \vert {\sqrt{f^{2}-\lambda^{2}}}\;{\rm s}\\
    i\vert \lambda \vert \;(1-{\rm c})&
    \lambda^{2}+f^{2} \;{\rm c}& 
    -f {\sqrt{f^{2}-\lambda^{2}}}\;s\\
    -i\vert \lambda \vert {\sqrt{f^{2}-\lambda^{2}}}\;{\rm s}&
    f {\sqrt{f^{2}-\lambda^{2}}}\;{\rm s}&
    (f^{2}-\lambda^{2})\;{\rm c}
      \end{array}    \right) \nonumber \\
&\times&
\frac{1}{f^{2}-\lambda^{2}} 
{\bf G}(z_{0},\zbar_{0})\;, \label{parallel1}
\eeq
where $\rm c$ and $\rm s$ are abbreviations for  
\beq
{\rm c}&=&\cos \left[ \frac{f^{2}-\lambda^{2}}{2\vert \lambda \vert^{3/2}}\;e^{i\theta}\;(z-z_{0})+
\frac{\vert f^{2}-\lambda^{2} \vert}{2\vert \lambda \vert^{3/2}}\;e^{-i\theta}\;
(\zbar-\zbar_{0}) \right] \;, \nonumber \\
{\rm s}&=&\sin \left[ \frac{f^{2}-\lambda^{2}}{2\vert \lambda \vert^{3/2}}\;e^{i\theta}\;(z-z_{0})+
\frac{\vert f^{2}-\lambda^{2} \vert}{2\vert \lambda \vert^{3/2}}\;e^{-i\theta}\;
(\zbar-\zbar_{0}) \right] \;. \nonumber
\eeq
While the expression \rf{parallel1} gives a real field strength $F^{a}$
only if $\lambda^{2}\le f^{2}$, we can
continue the solution to any $\lambda<0$.

The expression \rf{parallel1} is simple to use to find the form of the gauge field
and the field strength everywhere. If we choose $\theta=0$, so that the spatial
dependence is in the $x^{1}$-direction, the result is
\beq
A_{1}^{a}&=&\frac{f}{\vert \lambda\vert^{3/2}} \gamma_{0}^{a} \;, \nonumber\\
A_{2}^{a}&=&\frac{1}{\vert \lambda \vert^{1/2}}\beta_{0}^{a} \cos
\frac{(f^{2}-\lambda^{2})(x^{1}-x^{1}_{0})}{\vert\lambda\vert^{3/2}}
-\frac{1}{\vert \lambda \vert^{1/2}} F_{0}^{a}\sin\frac{(f^{2}-\lambda^{2})(x^{1}-x^{1}_{0})}{\vert \lambda\vert^{3/2}}
\;,\nonumber \\
F^{a}&=&\beta_{0}^{a} \sin\frac{(f^{2}-\lambda^{2})(x^{1}-x^{1}_{0})}{\vert \lambda\vert^{3/2}}
-F_{0}^{a}\cos\frac{(f^{2}-
\lambda^{2})(x^{1}-x^{1}_{0})}{\vert \lambda\vert^{3/2}}\;,
\nonumber
\eeq
where $F^{a}_{0}$, $\gamma^{a}_{0}$, $\beta^{a}_{0}$ are the field strength and the isovectors in 
\rf{realparam}, $F^{a}(z,\zbar)$, $\gamma^{a}(z,\zbar)$ and
$\beta^{a}(z,\zbar)$, respectively, evaluated at 
$z_{0}=x^{1}_{0}+ix^{2}_{0}$ and
$\zbar_{0}=x^{1}_{0}-ix^{2}_{0}$.

\vfill

\end{document}